\documentclass[aps,prb,reprint,superscriptaddress,showpacs,amsmath,amssymb,nofootinbib]{revtex4-2}

\usepackage[utf8]{inputenc}
\usepackage[T1]{fontenc}
\usepackage{amsmath}
\usepackage{amssymb}
\usepackage{MnSymbol}
\usepackage{bm}
\usepackage{graphicx}
\usepackage{hyperref}

\begin{document}

\title{Wave-Packet-Based Coherent Transport as a Time-Domain Dual to the Landauer-Büttiker Approach}

\author{Andrzej Biborski}
\email{andrzej.biborski@agh.edu.pl}
\affiliation{AGH University of Krakow, Academic Centre for Materials and Technology, al. A. Mickiewicza 30, 30-059 Krakow, Poland}

\begin{abstract}
Quantum transport in mesoscopic systems is conventionally formulated within the Landauer--Büttiker scattering framework, where steady-state currents emerge from the transmission of plane waves representing propagating carriers. While highly successful, this description obscures the explicit time-domain dynamics of individual fermionic excitations and their role in establishing macroscopic transport. Here, we present an exact and self-contained time-domain construction of Landauer transport based on a discrete basis of orthogonal fermionic wave packets. Starting from a second-quantized formulation, we define packet creation operators via a continuous Fourier transform over a finite transport energy window. By encoding  the Pauli exclusion principle, which enforces a fundamental temporal spacing $\Delta t = h/eV$, the current is reproduced in terms of orthogonal wave packets that are used for the rigorous construction of the many-body fermionic state. In this representation, a noiseless current emerges as a deterministic sequence of charge-carrying events, yielding the Landauer conductance $G_0 = e^2/h$. Additionally, the underlying Fock space decomposition into finite disjoint energy sub-bands renders the numerical approach  scalable for high performance computing platforms. Furthermore, the invoked packet language can be utilized to model the interacting carriers within truncated Fock space, constructed in a single particle basis of time-independent orthogonal wave packets characterized by a quantum number indicating the time of the discrete  emission event. Our results establish a direct and rigorous bridge between the continuous scattering description of quantum transport and a discrete, time-resolved picture based on fermionic wave packets. 

\end{abstract}

\maketitle

\section{Introduction}
Quantum transport in mesoscopic systems is commonly described within the Landauer--Büttiker (L--B) scattering formalism \cite{Landauer,Buttiker,Buttiker1990,Langreth}, which relates macroscopic observables, such as conductance and shot noise, to the transmission probabilities of extended scattering states. This approach is remarkably successful when the fermionic carriers can be treated as effectively independent, mutually non-interacting particles. When particle-particle interactions or strong electronic correlations dictate the system's macroscopic properties, advanced many-body techniques—most notably the Non-Equilibrium Green's Function (NEGF) formalism \cite{Caroli_1971,Jauho1994,Datta_1995,Croy,Ozaki2010,Sieberer_2016,Kloss_2021}, alongside emerging time-dependent tensor network methods—become indispensable\cite{Schmitteckert2004,White2004,Daley_2004,SCHOLLWOCK201196}. Nevertheless, for a vast class of fundamental transport phenomena, ranging from archetypal tunneling-based effects in semiconductor junctions \cite{Esaki1958} to the integer quantization of conductance in quantum point contacts \cite{vanWees}, the purely coherent, single-particle L--B framework remains  effective and fundamentally justified. This framework, rooted in the works of Landauer and Büttiker, provides a remarkably successful and computationally efficient description of phase-coherent transport\cite{Lent,Shao}. However, its standard formulation is intrinsically energy-domain based and therefore obscures the explicit spatio-temporal structure of the underlying quantum dynamics.

In parallel, the rapid development of electron quantum optics has established a complementary paradigm in which transport is described in terms of localized, time-resolved single-electron excitations\cite{bauerle2018coherent}. The realization of on-demand sources\cite{feve2007demand} and minimal-excitation \cite{dubois2013minimal} states has demonstrated that electronic transport can be naturally understood as a sequence of propagating fermionic wave-packets\cite{Hassler2008,Bocquillon,GAURY20141}. Within this picture, the Pauli exclusion principle enforces strong temporal correlations between emitted particles, leading to anti-bunching and the suppression of shot noise in ballistic conductors\cite{Martin,blanter2000shot,YAMAMOTO199719,Muller2010,bauerle2018coherent}.
Despite these advances, a fully explicit and analytically controlled connection between the continuous energy-domain L-B formalism and a discrete time-domain wave-packet description remains incomplete. Specifically, the exact formal mapping between these two regimes, as well as the pragmatic applicability of the time-domain approach to simulate noiseless macroscopic transport under DC and quasi-DC biases, requires a rigorous mathematical formulation. In this work, we present a self-contained framework that establishes an exact mapping between these two representations. Note that we do not intend to re-evaluate the well established L-B formalism, which handles transport in an elegant way, relying on the plane wave basis. We aim to provide  its formulation based on time-energy duality, which reveals the opportunity to use it directly in real-time dynamics. In this spirit, the presented approach exposes the full completeness of the L-B formalism for coherent transport of non-interacting fermions in this view, which is extendable to the interacting case.

In our formulation, we  use second quantization, which provides a compact way to represent algebraic structures and is particularly useful when interactions between fermions are taken into account; however, we also briefly present the mapping onto the regular first-quantization picture as it is used in the illustrative numerical experiments. 
By introducing packet creation operators obtained via a continuous Fourier transform of energy-resolved fermionic modes, the Pauli principle enforces a discrete temporal lattice with a fundamental spacing $\Delta t = h/eV$\cite{LANDAUER1991167,Martin}.  
Within this framework, we derive the Landauer conductance formula directly in the time domain, without invoking momentum-space kinematics or semiclassical arguments.

Both the unitary energy-time Fourier transform and the Pauli exclusion principle dictate the formulation of Landauer ballistic current in the picture of  collectively moving fermionic wave-packets. 

The paper is organized as follows: first, we provide the essential transformation back from the L-B approach to the packet based dual picture in the spirit of the works of Martin and Landauer\cite{Martin} and Hassler et al\cite{Hassler2008}. Next, we show that our formulation leads to the same conclusions as the standard L-B formalism does in view of steady state properties such as transmission and shot-noise.  Subsequently, by using this formalized language, we establish the packet-based approach suitable for situations where electronic correlations play an important or even crucial role. Eventually, we illustrate our analytical findings by presenting the numerical experiments performed explicitly in the time domain for both non-interacting and interacting cases. In the last section, we summarize the presented approach and discuss its applicability and possible extensions.

\section{Theoretical framework}

\subsection{Motivation: Time-Energy Duality}

The Landauer-Büttiker (L-B) formalism has long served as the cornerstone for describing coherent quantum transport in mesoscopic systems. However, its inherently stationary nature, rooted exclusively in the energy-momentum domain, naturally obscures the explicit time-resolved dynamics and the corpuscular nature of individual charge carriers. The primary motivation of the present framework is to establish a rigorous time-space-oriented approach that acts as a direct \textit{dual} to the established L-B picture. Rather than replacing the stationary approach, we aim to demonstrate that the continuous-wave scattering picture and the dynamic, wave-packet-based corpuscular picture are mathematically equivalent manifestations of the same underlying physical reality.

From a methodological perspective, this duality becomes particularly transparent when comparing our approach to standard numerical implementations of the L-B formalism, such as the Quantum Transmitting Boundary Method\cite{Lent}(QTBM). In the conventional QTBM framework, the physical model is constructed by applying a Fourier transform to the time-dependent Schrödinger equation, thereby mapping the system into the energy domain. Consequently, the influence of the macroscopic reservoirs is imposed via specific \textit{boundary conditions} in the form of continuous, monochromatic plane waves. This formulation establishes a classic Boundary Value Problem (BVP) that seamlessly yields the steady-state macroscopic Landauer current.

\begin{figure}
    \centering
    \includegraphics[width=0.99\linewidth]{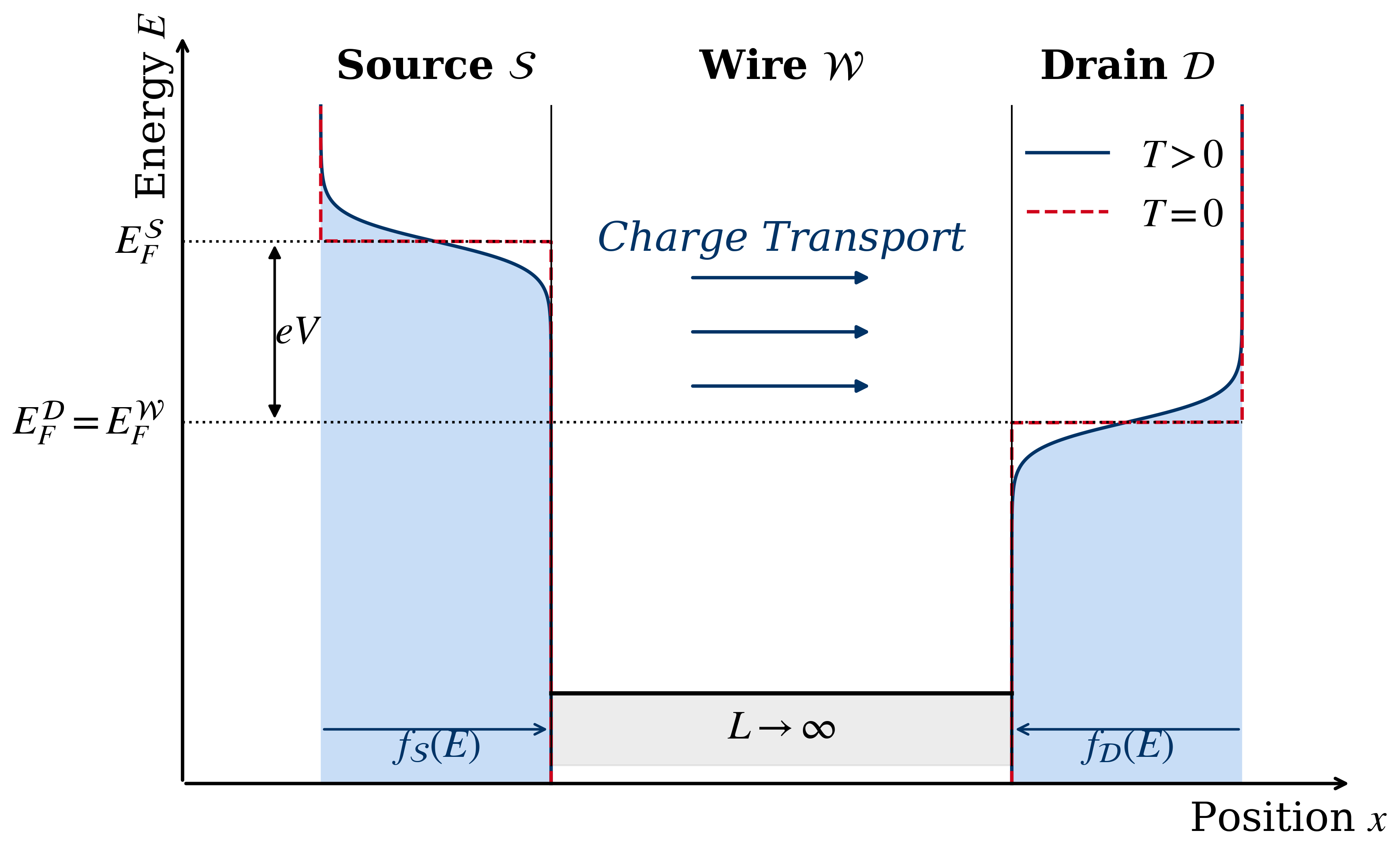}
    \caption{ Conceptual illustration of the wave-packet-based transport formalism. The standard Landauer approach, which relies on plane waves and boundary conditions in the energy window $(E_F, E_F+eV)$, is unitarily transformed into a time-domain equivalent. In this dual picture, the exact ballistic current $I=e^2V/h$ emerges from an initial many-body state constructed as a train of localized wave packets.}
    \label{fig:strategy}
\end{figure}

Conversely, if one forgoes the Fourier transform to explicitly investigate real-time dynamics using localized particles, the fundamental physical constraints imposed by the reservoirs and quantum statistics do not vanish. Instead, they must be mathematically translated. In this time-domain representation, the constraints shift from the spatial boundaries of the system directly onto the \textit{initial conditions} of the time-evolution process. The transport description fundamentally transforms from a Boundary Value Problem in the energy domain into an Initial Value Problem (IVP) in the time domain. The sketch of strategy applied by us is depicted in Fig.\ref{fig:strategy}.

To successfully realize the L-B formalism and reconstruct the exact macroscopic coherent current, the initial conditions cannot be chosen arbitrarily. The requirement to rigorously preserve the Pauli exclusion principle for a continuous stream of particles dictates  specific many-body state construction. Therefore, the choice of Wannier wave packets to describe the localized electrons, alongside the requirement for their emission at specific temporal intervals ($\Delta t = h / \Delta E$, where $\Delta E=eV$ when referring to Fig.\ref{fig:strategy}), is not an ad hoc modeling convenience. As will be derived in the following sections, they are the mathematically requisite initial conditions that dynamically and exactly realize the L-B scattering formalism within the real-time domain.

\begin{figure}
    \centering
    \includegraphics[width=0.99\linewidth]{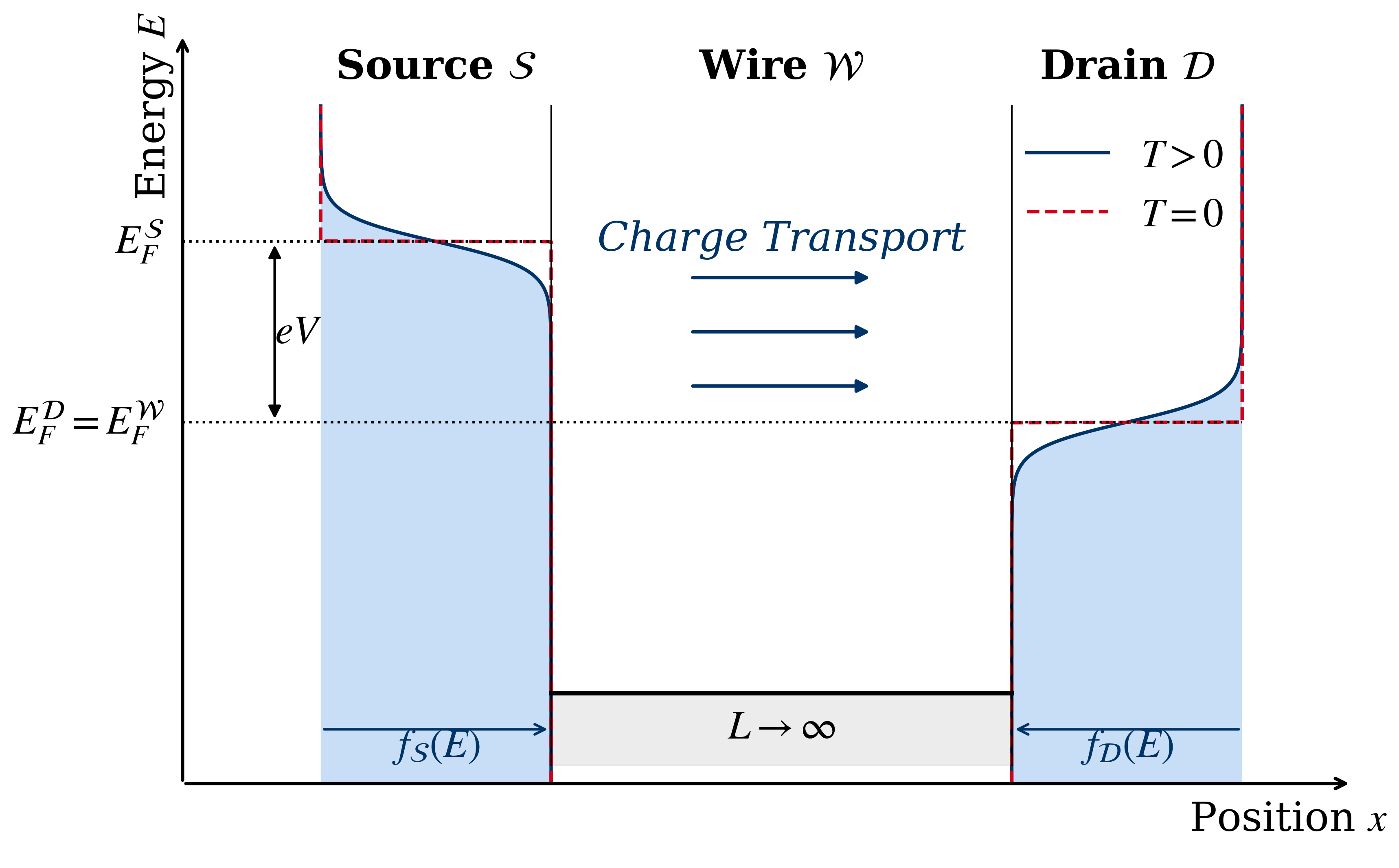}
    \caption{The sketch of the system considered here. }
    \label{fig:1}
\end{figure}

Our theoretical framework considers a one-dimensional channel of length $L$ through which a DC current flows.  Charge carriers are injected from a source reservoir $\mathcal{S}$ obeying standard Fermi-Dirac statistics. The application of a bias voltage $V$ between the source $\mathcal{S}$ and the drain $\mathcal{D}$ shifts the source chemical potential to $E_F^{\mathcal{S}} = eV$ (see Fig.\ref{fig:1}). Throughout the following derivations, we evaluate the system in the limit $L \rightarrow \infty$, which, in practical implementation, reduces to the condition that the size of the possibly immersed scattering center is small compared to $L$.

\subsection{Wave packets picture at $T=0$}

We start our consideration on the basis of the conception provided by Martin and Landauer\cite{Martin} and later by Hassler et al. \cite{Hassler2008}; however, in our formulation, we take advantage of the compactness of the second quantization framework. We emphasize that the creation and annihilation operators introduced below are employed as algebraic Fock-space tools to construct antisymmetric many-body states, perform unitary basis transformations, and enforce fermionic statistics, rather than describing high-energy particle creation or destruction. For the sake of brevity, we first implicitly disregard the spin degree of freedom, which can be trivially incorporated into the presented model, namely we consider this degree of freedom in the very last numerical illustration. 
To make the precise link between the physical situation depicted in Fig.\ref{fig:1} and the strategy provided in Fig.\ref{fig:strategy}, let us present the field operator in the form
\begin{equation}
    \hat{\Psi}(x,t)=\sum_{\zeta \in\{1,-1\}}\int_{\mathcal{E}}d\epsilon\phi_{k^{\zeta}_{\epsilon}}(x,t)\hat{c}
_{k^{\zeta}_{\epsilon}},
\end{equation}
where $\phi_{k^{\zeta}_{\epsilon}}(x,t)\sim e^{ik^{\zeta}_{\epsilon}x}$ is a plane wave basis characterized by the momentum $\zeta|k(\epsilon)|$, and $\zeta\in\{+1,-1\}$ defines the chirality (direction of propagation) of each wave, provided that $q^\zeta$ is uniquely determined by energy $\epsilon$. Operators $\hat{c}_{k^{\zeta}_{\epsilon}}$/$\hat{c}_{k^{\zeta}_{\epsilon}}^{\dagger}$ are fermionic annihilation/creation operators conforming to standard anticommutation rules
\begin{align}
    \{\hat{c}_{k},\hat{c}_{k'}^{\dagger}\} &= \delta(k-k'), \\
    \{\hat{c}_k^{},\hat{c}_{k'}^{}\} &= \{\hat{c}^{\dagger}_k,\hat{c}^{\dagger}_{k'}\} = 0.
\end{align}
For the sake of brevity, we assign the energy characterizing waves with $\zeta=1$ and momentum $|k(E_F^{\mathcal{D}}\leq\epsilon\leq E_F^{\mathcal{S}})|$ -- that is, those propagating from the energy window $eV$ -- by $E$. The states that do not appear to set this condition will be denoted by $q$. Then, one may formally decompose the Hilbert space $\mathcal{H}_{tot}$ spanned by plane waves into the direct sum $\mathcal{\mathcal{H}}_E\oplus\mathcal{H}_q$; we simply assign $\mathcal{H}\equiv\mathcal{H}_E$. If not stated explicitly, the algebra presented below is carried out for states $\hat{c}_E^{\dagger}|0\rangle\in\mathcal{H}$.

For the energy window of magnitude $eV$, which represents the spectrum of states participating in transport, and at $T=0$, \footnote{The $T > 0$ case can be considered approximately but in a controllable manner by decomposing the energy window $eV$ into finite slices (sub-bands) - as we discuss in sub-section II.E.2.} we define the \emph{packet creation} operator at a given emission time $t_n$ as:
\begin{align}
    \hat{A}_n^{\dagger} &= \frac{1}{\sqrt{eV}}\int_{E_f}^{E_f+eV}e^{iEt_n/\hbar}\hat{c}_E^{\dagger}dE, 
    \label{eq:centralunitary}
\end{align}
that is, for states in $\mathcal{H}$, we perform the unitary transformation from the energy to the time domain in a manner similar to that of Martin and Landauer \cite{Martin}, and Hassler et al. \cite{Hassler2008}. 
By evaluating the anticommutator for the packet operators, one finds that 
\begin{align}
    \{\hat{A}_m^{},\hat{A}_n^{\dagger}\} &= e^{ieV(t_n-t_m)/2\hbar}\dfrac{\sin[eV(t_n-t_m)/2\hbar]}{eV(t_n-t_m)/2\hbar} \nonumber \\
    &= e^{ieV(t_n-t_m)/2\hbar}\text{sinc}[eV(t_n-t_m)/2\hbar],
    \label{eq:packet_anti}
\end{align}
and,
\begin{align}
    \{\hat{A}_m^{\dagger},\hat{A}_n^{\dagger}\} &= \{\hat{A}_m^{},\hat{A}_n^{}\} = 0.
\end{align}
Nevertheless, operators $\hat{A}_{n}$ also preserve the canonical anti-commutation rules with the states $\hat{c}_q^{\dagger}|0\rangle \in\mathcal{H}_q$, as can be shown in an elementary manner.

The anticommutator given in Eq. (\ref{eq:packet_anti}) vanishes  when the time interval is quantized as 
\begin{align}
    \Delta t = |t_n - t_m| = h/eV,
    \label{eq:time_interval}
\end{align}
yielding a discrete, orthogonal basis $\{\hat{A}_n^{},\hat{A}_m^{\dagger}\} = \delta_{n,m}$. Thus, the packets are represented by independent fermionic \emph{quasiparticles} injected at regular time intervals\cite{Martin,Stevens_1987}. That is, when formulating L-B in the wave-packet-oriented approach, condition stated in Eq.(\ref{eq:time_interval}) guarantees that these quasiparticles obey the Pauli exclusion principle. Thus, the many body state referring to the formation of Landauer coherent current can be constructed from their single product, as will be proved.

To map the energy-domain quasiparticles onto the physical spatial coordinates of the single-channel wire, we formally evaluate the spatial Wannier functions by projecting the packet creation operators onto the real-space representation: $W_n(x,t) = \langle 0 | \hat{\Psi}(x,t) \hat{A}_n^{\dagger} | 0 \rangle$. 

Because the injected quasiparticle $\hat{A}_n^{\dagger}$ is constructed exclusively from the right-moving eigenmodes originating from the source reservoir, the vacuum expectation value identically annihilates all left-moving  and out-of-window components of the field operator. Consequently, the integral naturally collapses to the relevant right-moving positive energy domain. By performing a change of variables from momentum to energy for this surviving subspace, and ensuring the preservation of the anticommutation relations via the relation $\hat{c}_k^{} = \sqrt{\hbar v_g(E)} \hat{c}_E^{}$ (where $v_g(E) = \frac{1}{\hbar}\frac{\partial E}{\partial k}$ is the group velocity parameterized by energy due to monotonicity), the analytical form of the free-propagating Wannier function evaluates to:
\begin{align}
    W_n(x,t) = \frac{1}{\sqrt{2\pi \hbar eV}} \int_{0}^{eV} \frac{dE}{\sqrt{v_g(E)}} \, e^{i \left[ k(E) x - \frac{E}{\hbar} (t - t_n) \right]}.
    \label{eq:general_wannier}
\end{align}
This  representation highlights the role of the specific band structure. For an arbitrary non-linear dispersion $E(k)$ and an energy-dependent velocity $v_g(E)$, the integral in Eq. (\ref{eq:general_wannier}) is analytically convoluted, implying that the wave packet will generally suffer from spatial distortion (chirp) as it propagates.

\subsection{Landauer current: Ballistic transport of packets at $T=0$ }

 We demonstrate that the fundamental feature of Landauer transport—namely, the exact conductance quantization — is  protected by the temporal packing sequence, entirely independent of the underlying energy-momentum dispersion relation $E(k)$.

Assuming that the macroscopic transport window $[0, eV]$ does not encompass any local band extrema, the momentum $k(E)$ is a strictly monotonic function for the right-moving branch, we can utilize the transformation identity $\delta[k(E)-k(E')] = \hbar v_g(E) \delta(E-E')$ thus
\begin{align}
    \int_{-\infty}^{\infty} W_m^*(x,t) W_n(x,t) dx = \frac{1}{eV} \int_{0}^{eV} e^{iE(t_m - t_n)/\hbar} dE = \delta_{m,n},
\end{align}
proving that the unitary transformation from the energy domain to physical space  preserves the Pauli exclusion principle, regardless of the band structure complexity.

Let us consider the system  at zero temperature, where the voltage-biased source injects a continuous, fully occupied stream of these quasiparticles that propagate ballistically; that is, no scattering regions are present. The macroscopic many-body state of the channel is  represented by 
\begin{align}
|\Phi_0\rangle = \prod_{j=-\infty}^{\infty} \hat{A}_j^{\dagger} |0\rangle. 
\label{eq:train}
\end{align}

That is, it serves as an initial state in the IVP that describes the Landauer coherent current while conserving the Pauli exclusion principle.

To evaluate the macroscopic current, we consider the expectation value of current operator $\hat{I}(x,t)$ for the many-body state $|\Phi_0\rangle$. 
The local current operator is given as 
\begin{equation}
    \hat{I}(x,t)=\frac{e}{2}\bigg\{\hat{\Psi}^{\dagger}(x,t)\hat{v}_g\hat{\Psi}(x,t)+\big[\hat{v}_g\hat{\Psi}(x,t)\big]^{\dagger}\hat{\Psi}(x,t)\bigg\}    
\end{equation}
where $\hat{v}_g=\frac{1}{\hbar}\dfrac{\partial\hat{H}}{\partial\hat{k}}$ is the group velocity operator. We notice that after the provided transformations and evaluation of Wannier-like states (wave-packet functions), the field operator can be written in the form
\begin{equation}
    \hat{\Psi}(x,t)=\sum_{j=-\infty}^{\infty}W_j(x,t)\hat{A}_j+\int_{\mathcal{E}_q}d{\epsilon}_{q}\phi_{q}(x,t)\hat{c}_{q},
    \label{eq:fieldopertor_decomposed}
\end{equation}
as continuous spectrum of states from the transport energy window is unitarily transformed into the infinite series of packets $\hat{A}_j^{\dagger}|0\rangle$. Note that we use a convention in which $W_{j}(x,t)=\phi_j(x,t)$ is used interchangeably when the distinction between $\mathcal{H}$ and $\mathcal{H}_q$  is not obligatory.
By also using a single naming convention for the operators $\hat{A}_{j}$ and $\hat{c}_{q}$, along with $\hat{\alpha}_{\mu}$ for both (where the index $\mu$ possesses all the information needed to identify the state $\hat{\alpha}^{\dagger}|0\rangle$), one obtains the current by sum-integrating over the entire Hilbert space given as
\begin{equation}
\begin{split}
    I(x,t) &= \langle\Phi_0|\hat{I}(x,t)|\Phi_0\rangle =e \sumint I_{\mu\nu}(x,t) \langle\Phi_0|\hat{\alpha}_{\mu}^{\dagger}\hat{\alpha}_{\nu}|\Phi_0\rangle d\mu d\nu \\
    &= {e}\sum_{|n\rangle \in \mathcal{H}}\sum_{|m\rangle\in{\mathcal{H}}} I_{nm}(x,t) \langle\Phi_0|\hat{A}_n^{\dagger}\hat{A}_m|\Phi_0\rangle =  e\sum_{|n\rangle\in\mathcal{H}} I_{nn}(x,t),
    \label{eq:elegant_current}
\end{split}
\end{equation}
where matrix elements $I_{\mu\nu}(x,t)$ are defined as
\begin{equation}
      I_{\mu\nu}(x,t)=\frac{1}{2}\bigg\{\phi^{*}_{\mu}(x,t)\big[\hat{v}_g\phi^{}_{\nu}(x,t)\big]+\big[\hat{v}_g\phi_{\mu}(x,t)\big]^{*}\phi_{\nu}(x,t)\bigg\}.\label{eq:matrix_current}
\end{equation}
This result is somewhat natural since we intentionally assume that only states $\hat{A}^{\dagger}|0\rangle$  participate in transport.

Eventually, by combining Eqs.(\ref{eq:elegant_current},\ref{eq:matrix_current}) with Eq.(\ref{eq:general_wannier}), we are left with 
\begin{equation}
\begin{aligned}
   I(x,t)  &= \frac{e}{eV} \sum_{n=-\infty}^{\infty} \int_{0}^{eV} \int_{0}^{eV} dE dE' \, \frac{v_g(E)+v_g(E')}{2h \sqrt{v_g(E)v_g(E')}} \, \times \\
    &\quad \times e^{-i[k(E)-k(E')]x} \, e^{i(E-E')(t-t_n)/\hbar}.
    \label{eq:current}
\end{aligned}
\end{equation}
Remarkably, by extracting the sum over the temporal indices $t_n = n(h/eV)$, we encounter the Poisson summation formula, which yields a Dirac comb in the energy domain:
\begin{align}
    \sum_{n=-\infty}^{\infty} e^{-i(E-E') n \frac{h}{eV\hbar}} = eV \sum_{j=-\infty}^{\infty} \delta(E-E' - j\cdot eV).
\end{align}
Because both integration variables $E$ and $E'$ are  bounded within $[0, eV]$, the only term that contributes to the integral is the central peak at $j=0$, reducing the comb to $eV \delta(E-E')$. Applying this exact constraint to the total current equation instantly annihilates all spatio-temporal phase fluctuations ($k(E)-k(E') \rightarrow 0$) and enforces $v_g(E') \rightarrow v_g(E)$. The group velocity originating from the current operator exactly cancels the velocity-dependent normalization factor of the injected packets, reducing the macroscopic steady-state flow to:
\begin{equation}
    \langle I(x,t) \rangle = e \int_{0}^{eV} dE \, \left( \frac{v_g(E)}{h v_g(E)} \right) = \frac{e^2}{h} V.
\end{equation}

This result proves that the assumed initial form of the many-body state, formulated within elaborated Wannier states $W_n(x,t)$, is fully equivalent to the standard L-B approach formulated in terms of the plane wave basis. Obviously, this  wave-packet picture is not limited to non-scattered ballistic transport (i.e., no scattering on barriers, wells, etc.), nor is it limited only to the $T=0$ in the practical implementation; we clarify this statement in the following subsections.

\subsection{Beyond non-scattering case: transmission coefficient, current-current correlations and shot noise}

Up to now, we have shown how to construct the Landauer current using a many-body state consisting entirely of localized wave packets represented by time-dependent Wannier functions $W_n(x,t)$. A natural question arises: how to incorporate a general scattering potential $V(x)$ and extract transport properties, namely the transmission and shot noise. In this section, we demonstrate how this is achieved within our packet-oriented Landauer-B\"uttiker (L-B) formalism. 
\begin{figure}
    \centering
    \includegraphics[width=0.95\linewidth]{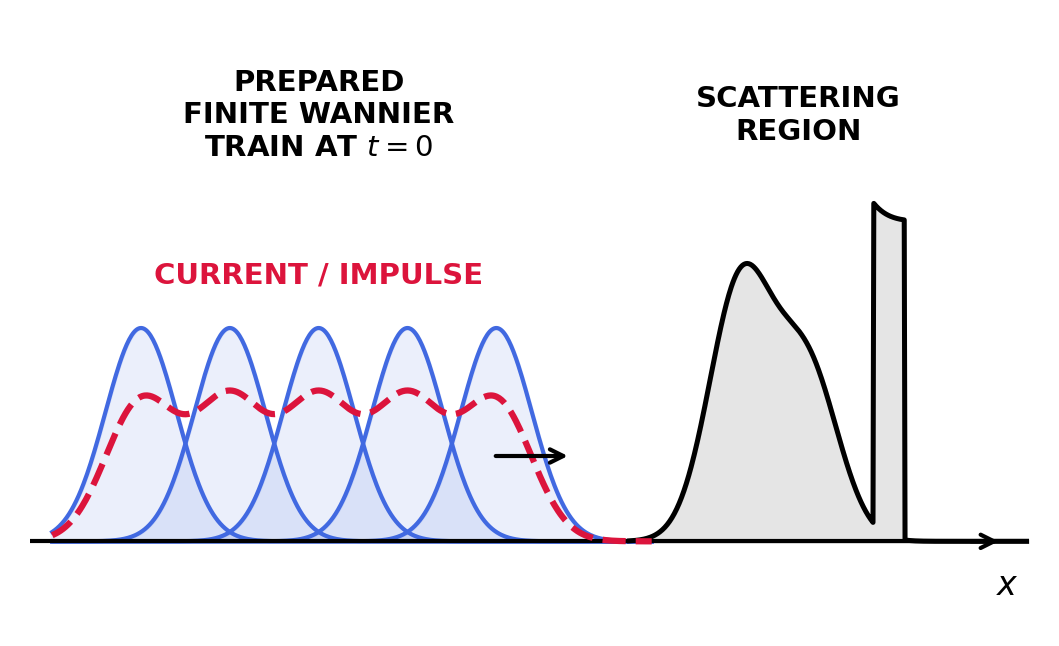}
    \caption{The illustration demonstrating how the transport with scattering potential can be realized in the language of orthogonal packets by providing finite Wannier train (blue areas) which at infinite number packets reconstruct Landauer current (red dashed line). The increase in the number of packets reduces artificial oscillations as shown in exemplary numerical simulations presented devoted Section. Note, that any form of potential $V(x)$ represented by grayed area can be considered in practical simulations.  }
    \label{fig:scattering}
\end{figure}
First, while the theoretical steady-state requires an infinite train, any practical simulation of transient transport---or a voltage pulse---must be limited to a finite series of $N$ Wannier states. These packets are injected at regular time intervals $\Delta t$ into the scattering region governed by an arbitrary potential profile $V(x)$. We consider this finite $N$-body state as the frozen \emph{initial} state $|\Phi_0\rangle$, which asymptotically recovers the continuous Landauer steady-state current in the limit of $N \to \infty$. The reasoning concerning the physical situation described is  provided in an illustrative manner in Fig.\ref{fig:scattering}.

Let the system be governed by a Hamiltonian $\hat{H}=\hat{T}+\hat{V}(x)$, where $\hat{T}$ is the kinetic term dictated by the assumed dispersion relation. Working in the Heisenberg picture, the initial many-body state $|\Phi_0\rangle$ remains time-independent. The time dependence is instead carried entirely by the field operator, whose single-particle basis functions—both the occupied Wannier packets $W_j(x,t)$ and the unoccupied states $\phi_q(x,t)$—evolve unitarily according to $i\hbar \partial_t \phi(x,t) = \hat{H} \phi(x,t)$. 

Crucially, because the scattering potential $V(x)$ breaks translational symmetry, the physical momentum is no longer a conserved quantum number. Therefore, while the field operator retains the decomposed form of Eq.~\ref{eq:fieldopertor_decomposed} (with the first summation now  limited to the finite set of injected packets), the index $q$ of the unoccupied states—and analogously, the initial momenta of the Wannier packets—ceases to represent the physical momentum at $t>0$. Instead, these indices serve strictly as labels, uniquely identifying the states as those that were either injected or empty in the distant past. This guarantees the preservation of orthogonality between the occupied and unoccupied subspaces throughout the entire time evolution, allowing for the treatment of the scattering process.

\subsubsection{Transmission coefficient}
As first we show how the transmission coefficient for a given energy $E$ in energy window of width $eV$ can be obtained from packet-oriented approach.
To achieve this, we implement a virtual Time-Flux detector at a fixed spatial coordinate $x_D$ within the free-propagation region, which is beyond the scattering region.

The spectral amplitude of the transmitted signal in the energy domain is obtained by performing a temporal Fourier transform of the time-evolving Wannier wave function sampled at the detector site:
\begin{equation}
    A(E) = \mathcal{F}_t \{ W_0(x_D, t) \}(E) = \int_{-\infty}^{\infty} W_0(x_D, t) e^{iEt/\hbar} \, dt.
\end{equation}

Concurrently, the spectral profile of the incoming state injected at $t=0$ is evaluated via a spatial Fourier transform of the initial Wannier packet $W_0(x, 0)$ (defined in Eq. 19):
\begin{equation}
    \phi(k_E) = \mathcal{F}_x \{ W_0(x, 0) \}(k_E) = \int_{-\infty}^{\infty} W_0(x, 0) e^{-ik_E x} \, dx,
\end{equation}
where $k_E$ represents the momentum uniquely associated with the injection energy $E(k)$.

By  balancing the incoming and outgoing probability currents across the device, the energy-resolved stationary transmission coefficient $\mathcal{T}(E)$ takes the following  form:
\begin{equation}
    \mathcal{T}(E) = \left[ v_g(E) \right]^2 \frac{\left| \mathcal{F}_t \{ W_0(x_D, t) \}(E) \right|^2}{\left| \mathcal{F}_x \{ W_0(x, 0) \}(k_E) \right|^2}.
\end{equation}

\subsubsection{Current-current correlations}

Not only $\mathcal{T}$ , but also current-current correlation functions are crucial to characterizing the transport. 
To evaluate correlation functions 
\begin{equation}
    S(x,t,x',t')\equiv\langle\Phi_0|\hat{\Delta I}(x',t')\hat{\Delta I}(x,t)|\Phi_0\rangle,
    \label{eq:sxt}
\end{equation}
we use  Eq.\ref{eq:elegant_current}, where $\hat{\Delta I}(x,t)\equiv\hat{I}(x,t)-I(x,t)$. By applying Wick's theorem\cite{FetterWalecka} (see Appendix \ref{app:wick_shot_noise}), one obtains the general formula:
\begin{equation}
    S(x,t,x',t')=e^2\sum_{|n\rangle\in\mathcal{H}}\int_{|q\rangle\in\mathcal{H}_{q}}I_{nq}(x,t)I_{qn}(x',t')dq,
    \label{eq:sxtWick}
\end{equation}
Note that the equation above involves any combination of states from $\mathcal{H}_{tot}$ in $I_{\mu\nu}(x,t)$, including those represented by Wannier packets and plane waves as well.
The direct time evolution of all states (packets and waves representing empty states), even in a truncated Hilbert space, is cumbersome from a computational perspective. Namely, to achieve this, one is forced to truncate Hilbert space $\mathcal{H}_q$ and make it discrete to appropriately approximate the continuous part of the spectrum. While this kind of modeling is, in principle, possible, we do not explore this path here.
Nevertheless, this mixing issue is not present when one is interested in calculating equal-time correlation functions $S(x,x',t)$ and the shot noise $S=\int dt\int dt' S(x_D,t,x_D,t')$.

Namely, when $x\neq x'$, the non-diagonal equal-time correlation function $S(x,x',t)$ is given by 
\begin{equation}
    S(x,x',t)=-e^2\sum_{|n\rangle\in\mathcal{H}}\sum_{|m\rangle\in\mathcal{H}}I_{nm}(x,t)I_{mn}(x',t),
    \label{eq:correlations_non_local}
\end{equation}
that is,  it does not mix occupied states in $\mathcal{H}$ with the continuous part of the whole Hilbert space (see Appendix \ref{app:wick_shot_noise} for details).

\subsubsection{Shot noise}

Similarly, the shot noise defined as $S(x_D)=\int dt\int dt'S(x_D,t,x_D,t')$ can be evaluated by integrating over $t$ and $t'$, and also by using the completeness relation $\sum_{q\in\mathcal{H}_q}=\mathbf{I}-\sum_{n\in{\mathcal{H}}}$. This leads to 
\begin{equation}
S(x_D)=e^2\sum_{|n\rangle\in\mathcal{H}}\bigg(P_{nn}(x_D)-\sum_{|m\rangle\in\mathcal{H}}|P_{nm}(x_D)|^{2}\bigg),
\label{eq:shot_noise}
\end{equation}
where 
\begin{equation}
    P_{\mu\nu}(x_D)=\int_{-\infty}^{\infty}I_{\mu\nu}(x_D,t)dt=\\\int_{x_D}^{\infty}\lim_{t\rightarrow\infty}\big[\phi_{\mu}^{*}(x,t)\phi_{\nu}(x,t)\big]dx.
    \label{eq:pmunu}
\end{equation}
Let us briefly discuss the formula \ref{eq:shot_noise} to refer it to the standard L-B result at $T=0$; that is, $S\propto\int dE\mathcal{T}(E)[1-\mathcal{T}(E)]$ \cite{blanter2000shot}. By evaluating $P_{nn}(x_D)$ in such a way that diagonal elements $I_{nn}(x_D)$ are assessed with the help of Wannier packets as defined in Eq.\ref{eq:general_wannier}, one obtains $P_{nn}(x_D)\propto\int_{-\infty}^{\infty} I(x_D,E)dE$. Namely, the application of Wannier functions in  Eq.\ref{eq:pmunu} is effectively the Fourier transform of $I_{nn}(x_D,t)$ into the energy domain; that is, $I(x_D,E)=\mathcal{F}_t\{I_{nn}(x_D,t)\}(E)$. Since $I(x_D,E)\propto\mathcal{T}(E)$, one finds that $P_{nn}(x_D)\propto\int_{eV}\mathcal{T}(E)dE\propto\langle\mathcal{T}(E)\rangle_{eV}$. Similar reasoning leads to the conclusion that $\sum_{|m\rangle}|P_{nm}|^2\propto\int_{eV}\mathcal{T}^2(E)dE\propto\langle\mathcal{T}^2\rangle_{eV}$. Eventually, provided $x_D$ is selected properly, one finds that an averaged value of the shot noise with respect to the averaged number of emitted packets  $\langle N \rangle_{\tau}\equiv\frac{\tau}{\Delta t}=\frac{\tau eV}{h}$ during the time interval $\tau$  is
\begin{equation}
    s_{\tau}=\langle S(x_D)\rangle_{\tau}\propto\int_{eV} \mathcal{T}(E)[1
-\mathcal{T}(E)]dE=\int_{eV}s(E)dE.
\end{equation}
The above result illustrates again, analytically, that the packet-based approach, as presented, is fully consistent with the standard L-B formalism elaborated within the plane-wave basis. This is, however, expected since the packets are constructed within the unitary transformation.
The numerical evidence of this agreement will be presented in the section devoted solely  to the demonstration of simulations carried out for exemplary physical scenarios.

\subsection{Invariance, Finite Temperatures and the Algebra of Orthogonal Wannier Trains}

Although the packet framework, which we have presented, is complete with regard to $T=0$, it can be extended in practical implementation to $T>0$. Simultaneously, the idea behind this extension can serve as just a speedup in calculations, that is, to perform them in a massively parallel architecture. In this subsection, we discuss these aspects.
Accommodating the finite temperature ($T>0$) regime requires incorporating the continuous, non-rectangular shape of the Fermi-Dirac statistics $f(E)$. Within our time-domain framework, this is achieved rigorously without sacrificing the wave-packet corpuscular picture by exploiting the fundamental orthogonality of states originating from non-overlapping energy domains.

\subsubsection{Slicing in energy window}
Let the total energy axis (comprising both the fully occupied core and the thermal tail) be partitioned into a set of disjoint energy windows $\{ \Delta{E}_j \}$, where $\Delta{E}_j = [E_{j-1}, E_j]$.  At the single-particle level, this partitioning corresponds to decomposing the total Hilbert space $\mathcal{H}$ into an orthogonal direct sum of continuous energy subspaces: $\mathcal{H} = \bigoplus_j \mathcal{H}_j$. That is, \textit{slicing} the total energy window into $\Delta{E}_j$ intervals does not change the underlying physical assumptions; it is just a unitary decomposition of the same Hilbert space. However, it allows one to speed-up the numerical calculations and, more importantly, to simulate time varying currents and/or scenarios where the system is characterized by a non-zero temperature.

Crucially, each window dictates its own intrinsic temporal lattice spacing for injection, $\Delta t_j = h/\Delta E_j$. We define the generalized packet creation operator for the $j$-th window at its $n$-th temporal bin as:
\begin{equation}
    \hat{A}_{n,j}^{\dagger} = \frac{1}{\sqrt{\Delta E_j}} \int_{\Delta{E}_j} e^{i E t_n^{(j)} / \hbar} \hat{c}_E^{\dagger} dE,
\end{equation}
where $t_n^{(j)} = n \Delta t_j$. Therefore, the corresponding Wannier state $|W_{n,j}(t)\rangle$ refers to the Wannier function $W_{n,j}(x,t) = \langle 0 | \hat{\Psi}(x,t) \hat{A}_{n,j}^{\dagger} | 0 \rangle$. Because the energy sub-bands are  disjoint ($\Delta{E}_j \cap \Delta{E}_{j'} = \emptyset$ for $j \neq j'$), the corresponding creation and annihilation operators satisfy the anticommutation relations $\{ \hat{A}_{n,j}^{}, \hat{A}_{n',j'}^{\dagger} \} = \delta_{n,n'} \delta_{j,j'}$.

This exact orthogonality allows us to define a macroscopic \emph{Wannier train} operator for each artificially invoked sub-band, representing a coherent sequence of $K_j$ charge carriers flowing through the given energy window over a total simulation time $t_{tot}$:
\begin{equation}
    \hat{\mathcal{T}}_j^{\dagger} = \prod_{n=1}^{K_j} \hat{A}_{n,j}^{\dagger},
\end{equation}
where $K_j = \lfloor t_{tot} / \Delta t_j \rfloor$. Since the individual packet operators anticommute across different windows, these many-body train operators describe completely independent, non-interfering quasiparticle streams. Assuming the system's total Hamiltonian $\hat{\mathcal{H}}$ does not depend on time and is completely elastic (i.e., it does not induce transitions between different energy windows, $[\hat{\mathcal{H}}, \hat{P}_j] = 0$, where $\hat{P}_j$ is the projector onto $\mathcal{H}_j$), the entire macroscopic many-body state emerges as a strict tensor product of these independent macroscopic quasi-particles acting on the global vacuum:
\begin{equation}
    |\Psi\rangle = \left( \prod_j \hat{\mathcal{T}}_j^{\dagger} \right) | 0 \rangle.
\end{equation}

This decomposition reveals that the slicing of the energy manifold into pieces of magnitude $\Delta E_j$ reflects an exact structural isomorphism of Fock spaces. Because the underlying continuous single-particle Hilbert space is partitioned as a direct sum of orthogonal subspaces, the exponential property of the Fock space $\mathcal{F}$ ensures that the global many-body space is rigorously reconstructed via the graded (antisymmetrized) tensor product based on the exponential isomorphism $\mathcal{F}=\mathcal{F}(\bigoplus_j \mathcal{H}_j) \cong \bigotimes_j \mathcal{F}_{j}$, where $\mathcal{F}_{j}\equiv\mathcal{F}(\mathcal{H}_j)$. Consequently, as long as we consider the $T=0$ pure state, the physical observables (such as the Landauer DC) are  invariant under any arbitrary refinement of this energy partitioning. The tensor product mathematically guarantees that different grid choices perfectly map to the identical continuous macroscopic state.
In other words, the partitioning of $\Delta E $ into disjoint $\Delta E_j$ slices is just  the unitary transformation at $T=0$, as can be shown in an elementary manner (see Appendix \ref{app:slicing} for derivations). Thus, in turn, the observables are invariant under this transformation.

\subsubsection{Application to finite temperatures}

At finite temperatures ($T>0$), this invariance is formally lost because the Fermi-Dirac occupation profile is no longer a binary step function. That is, one cannot just plug Fermi-Dirac statistics into the definition of the packet creation operator as defined in Eq.\ref{eq:centralunitary}, since it would violate the anticommutation relations and, in turn, the whole construction. However, in the scenario where $T>0$, the energy-window decomposition turns from being an exact basis choice into a systematic \emph{coarse-graining scheme}, allowing us to treat the non-zero temperature in an approximate, albeit controllable, manner. Assuming that the single-particle observables of interest are diagonal in the basis of our independent orthogonal trains, the macroscopic Non-Equilibrium Steady State (NESS) can be formally described by a statistical density matrix $\hat{\rho}$. Consequently, the macroscopic expectation value of any time-dependent single-particle observable $\hat{O}(t)$ evaluates to the trace over this density matrix. This rigorously reduces the problem to a thermodynamically weighted linear combination of independent, time-dependent pure-state train responses:
\begin{equation}
    \langle \hat{O}(t) \rangle = \text{Tr}[\hat{\rho}\hat{O}(t)] \approx \sum_j f({E}_j) \langle 0 | \hat{\mathcal{T}}_j^{} \, \hat{O}(t) \, \hat{\mathcal{T}}_j^{\dagger} | 0 \rangle,
\end{equation}
where $f(E)=1/[e^{\beta (E-\mu)}+1]$ is a standard Fermi-Dirac probability distribution function. Thus, the observable $\langle \hat {O}(t)\rangle$(e.g., transmission coefficient) is obtained by weighting the outcome of each channel $\Delta E_j$ by the Fermi-Dirac distribution at a given temperature $T$.  This structural factorization holds immense pragmatic implications for numerical studies of transport. Unlike standard time-dependent Non-Equilibrium Green's Functions (NEGF) methods, which inherently scale as $\mathcal{O}(N_t^3)$ in their standard formulation\cite{Zhu_2025} due to memory integrals over the full simulation history, the evaluation of the steady state via orthogonal Wannier trains scales  linearly, $\mathcal{O}(N_t)$. Because the total time-evolution operator perfectly factorizes as $\hat{U}(t) = \prod_j \hat{U}_j(t)$, the total thermal response can be obtained by independently propagating packets from distinct sub-bands and directly superimposing their resultant steady-state charge or current densities. This renders the presented framework exceptionally well-suited for massively parallel high-performance computing (HPC).

Note that when considering this parallel paradigm, one must ensure that the total time of the sub-process $t_{tot}$ is assumed to be identical across all independent channels $\Delta{E}_j$. Since a non-uniform grid is employed in the sub-band division, the resulting number of packets $K_j$ emitted during simulations is inherently window-dependent. This is physically imposed by the quantization of the emission time $\Delta t_j = h/\Delta E_j$. Therefore, the total number of packets propagated across the entire simulation is $\mathcal{N}_{tot} = t_{tot} \sum_j 1/\Delta t_j$, ensuring that the continuous wave-packet trains switch off simultaneously across all energy channels.

\subsection{Time-dependent injection and unidirectional time-dependent currents}

Introducing a time-dependent potential within the scattering region, or considering inelastic multi-channel transport, breaks the commutation relation $[\mathcal{\hat{H}}, \hat{P}_j] = 0$. Consequently, the strict separability of the evolution for each energy window is lost. While the purely fermionic character of the multiparticle state is technically preserved in such a scenario, one can no longer exploit the advantageous Fock space decomposition $\bigotimes_j \mathcal{F}_j$. In turn, the numerical implementation would require the simultaneous preparation and coherent evolution of the entire broadband packet sequence, precluding the massive parallelization scheme and severely diminishing the computational elegance of the approach.

Conversely, the formalism naturally accommodates time-dependent injection from the reservoirs, which enables the simulation of unidirectional time-dependent currents. This is achieved by adiabatically modulating the chemical potential in the source ($\mathcal{S}$) lead, namely $\mu_{\mathcal{S}}(t) = E^{\mathcal{S}}_F + e\tilde{V}(t)$. In our approach, the fundamental basis for the transport dynamics is formed by unmodulated, continuous wave-packet trains at each allowed energy $E_j$. The creation operator for such a pure-state train is straightforwardly defined without any statistical weights by constructing each train along the recipe
$\hat{\mathcal{T}}_j^\dagger = \prod_{n=-\infty}^{\infty} \hat{A}_{n,j}^\dagger$.

The time-dependent occupation statistics and the charge superselection rules are not imposed at the level of these pure states. Instead, they are rigorously enforced during the calculation of macroscopic observables via the statistical ensemble trace. Generalizing the steady-state density matrix formalism to the time-dependent injection regime, the macroscopic expectation value $\langle \hat{O}(t) \rangle$ cannot be evaluated by assigning a single static thermal weight to the entire sequence $\hat{\mathcal{T}}_j^\dagger$. Because the individual wave-packets $\hat{A}_{n,j}^{\dagger}$ constituting the trains are  orthogonal in the time-domain, the trace of any single-particle observable decomposes into a sum over all energy windows $j$ and all discrete emission times $t_n$. The total macroscopic response is thus obtained by weighting the pure-state dynamical contribution of each individual wave-packet with the momentary statistical distribution of the macroscopic reservoir:
\begin{equation}
    \langle \hat{O}(t) \rangle = \text{Tr}[\hat{\rho}(t)\hat{O}(t)] \approx \sum_j \sum_{n=-\infty}^{\infty} f_{n,j} \langle 0 | \hat{A}_{n,j}^{} \, \hat{O}(t) \, \hat{A}_{n,j}^{\dagger} | 0 \rangle.
\end{equation}
Consequently, under time-dependent statistical mixtures, the macroscopic Wannier train conceptually decomposes back into its fundamental wave-packet basis. Because the macroscopic quantities of interest (such as local charge density or current) are  single-particle operators, their expectation values evaluated over the many-body state reduce entirely to single-particle matrix elements.

Specifically, the occupation weight $f_{n,j}$ for the $n$-th wave-packet emitted at time $t_n$ is determined by evaluating the Fermi-Dirac distribution at the instantaneous chemical potential $\mu_{\mathcal{S}}(t_n)$:
\begin{equation}    
    f_{n,j} = \left[ 1 + \exp\left(\frac{E_{\text{max}, j} - \mu_{\mathcal{S}}(t_n)}{k_B T}\right) \right]^{-1}.
\end{equation}
In the absolute zero-temperature limit ($T \to 0$~K), this continuous statistical weighting naturally reduces to a binary occupation mask (a Heaviside step function), effectively ``punching discrete holes'' in the emitted sequence based on the instantaneous bias. For finite temperatures ($T > 0$), it provides a continuous dynamic weighting that seamlessly captures thermally assisted AC transport without violating the underlying fermionic basis.

Finally, we note a fundamental trade-off governed by the Nyquist-Shannon sampling theorem. The temporal separation between subsequent packets in a given window is $\Delta t_j = h / \Delta E_j$. Consequently, the maximum angular frequency $\omega_M$ of the driving signal $\tilde{V}(t)$ that can be alias-free resolved within an isolated channel is bounded by the sampling rate, yielding $\omega_M < \pi \Delta E_j / \hbar$. Therefore, narrowing the energy bins $\Delta E_j$ improves the steady-state spectral resolution but inherently restricts the maximum frequency of the AC transients that the discrete Wannier basis provided in the \emph{finite computational box} can capture.
\subsection{Wannier train as Slater determinant}
To provide an intuitive first-quantization perspective on the many-body state constructed from the Wannier wave-packet train, which can subsequently be applied in real-time simulations, it is  instructive to map our second-quantized pure state back to a spatial representation. 

The state $|\Phi_0\rangle$ is defined via creation operators as an infinite product over all emission events, spanning from the infinite past to the infinite future, as stated in Eq.~(\ref{eq:train}). In the first-quantization picture, this corresponds  to an infinite Slater determinant. However, because the time-evolved Wannier packets $W_n(x,t)$ are spatially localized, any local physical observable (such as the current measured at a specific position within a finite time window) only has a non-zero overlap with a finite subset of these packets. Wave packets emitted far in the past have already traversed the active region, and those scheduled for future emission yield zero amplitude at the detector.

Consequently, for any practical evaluation or finite time-domain simulation, the relevant non-zero minor of this infinite matrix rigorously reduces to an effective, finite Slater determinant of dimension $M \times M$, where $M$ encompasses the active single-particle events in the system:
\begin{equation}
\begin{split}
    &\Phi(\{x_i\}, t) \equiv \frac{1}{\sqrt{M!}} \langle 0 | \hat{\Psi}(x_M, t) \dots \hat{\Psi}(x_1, t) | \Phi_0 \rangle \\
    &\quad = \frac{1}{\sqrt{M!}} \det \begin{bmatrix}
    W_{k}(x_1, t) & \dots & W_{k}(x_M, t) \\
    \vdots & \ddots & \vdots \\
    W_{k+M-1}(x_1, t) & \dots & W_{k+M-1}(x_M, t)
    \end{bmatrix},
\end{split}
\label{eq:slater}
\end{equation}
where $\{x_i\}$ denotes the coordinates of the $M$ currently interfering electrons, and the index $k$ shifts as the temporal window advances. Note that the construction of this dynamic many-body state becomes computationally feasible precisely due to the discrete, quantized nature of the wave-packet emissions. We notice that, although the many body wave-function given in Eq.\ref{eq:slater} may appear \emph{intuitive} , the reasoning presented in the above sub-sections provides rigorous argumentation for its forms. Also, it specifies \emph{how to construct} single particle "orbitals" $W_n(x,t)$ to tailor the initial state describing the coherent Landauer current in the packets language.

From a practical point of view, one introduces an artificial spatial boundary for the packet injections and numerically evaluates the time evolution of each Wannier function independently. This independent single-particle propagation remains valid, provided that many-body charge-charge interactions are neglected. Nevertheless, the time evolution of each wave-packet  is not the only possible approach, albeit effective for the non-interacting case. Yet another possibility relies on the time-independent Wannier functions, which is another option as we present in the section devoted to the transport when fermions interact with each other.
\subsection{Sketch of simulation algorithm}
In this part, we focus on the scenario when carriers do not interact and the whole dynamics is encoded in time evolution carried out for each considered Wannier packet.

\begin{figure}
    \centering
    \includegraphics[width=0.99\linewidth]{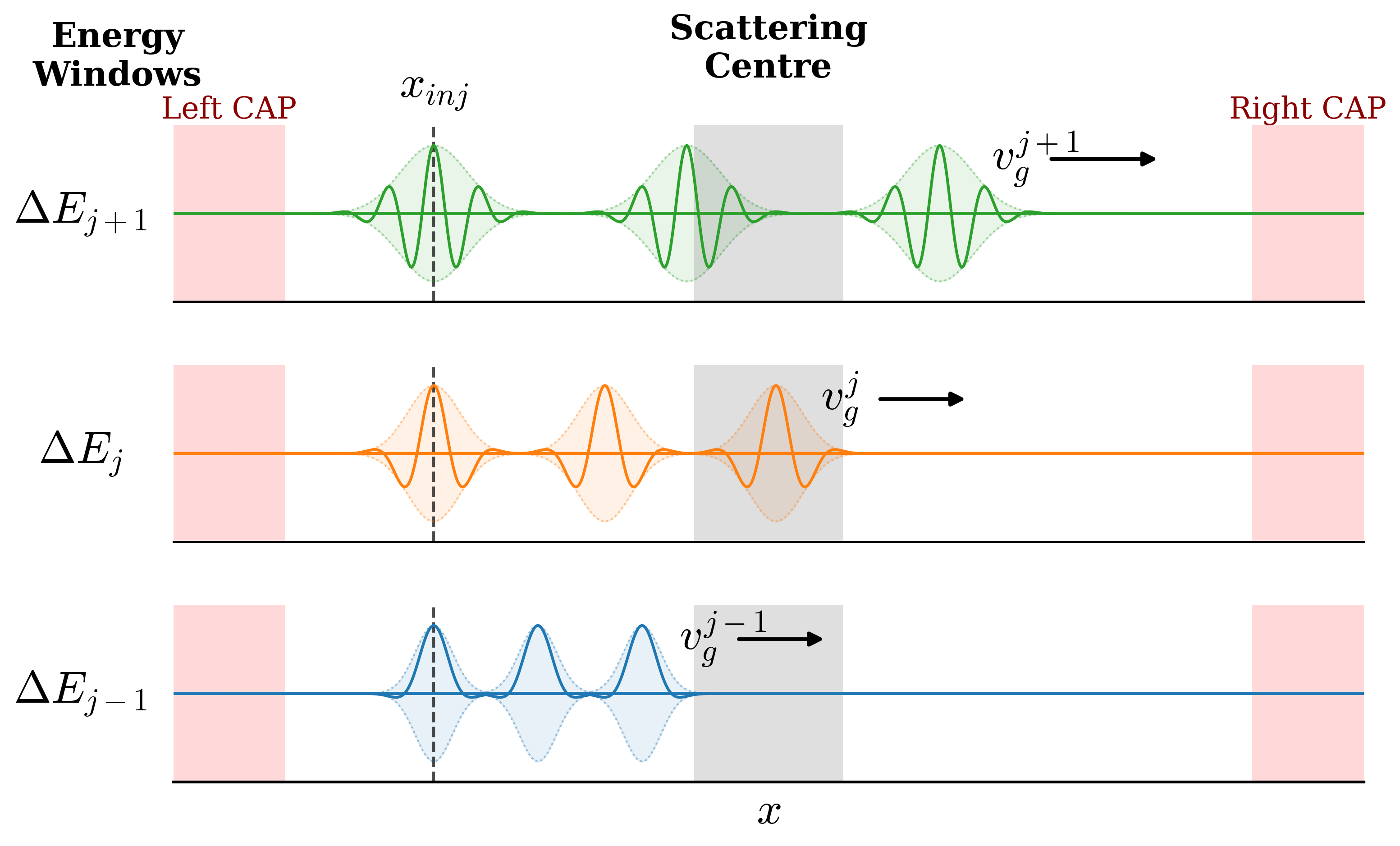}
    \caption{The cartoon  illustrating the simulation algorithm and its main assumptions. The simulation box is supplied with CAP on each boundary and optionally the  with scattering center. The packet propagation can be performed independently for each energy window $\Delta E_j$. Note, that group velocity $v_g$, as well as, the shape of packet and its oscillations generally depend on $\Delta E_j$, that is, not only on its width but also on  $E_{j/j-1}$. }
    \label{fig:algorithm_sketch}
\end{figure}

Obviously, the simulation has to be performed in the finite simulation box of length $L$ and simulation time $t_{sim}$. Assuming those are established, one then divides the transport energy window  into the set of $\{\Delta E_j\}$ channels in a way that is suitable for the given $T$ and possible source voltage time dependence. If $T=0$ and the incident current does not depend on time, or if the problem is not computationally complex, one may simply consider a single interval $\Delta E=eV$. The simulation box should be supplied with \emph{a Complex Absorbing Potential} (CAP) to limit the undesired scattering effects caused by the boundaries. The scattering center(s) (barriers, quantum wells) are also provided before the simulation starts, as shown in Fig.\ref{fig:algorithm_sketch}. Instead of propagating the full set of orthogonal packets prepared, one \emph{injects} packets at selected $x_{inj}$ with the time intervals $\Delta t_j$ in each channel. 
The temporal evolution of  $\Phi(\{x_i\},t)$ is computed by solving the time-dependent Schrödinger equation using the Crank-Nicolson\cite{Crank_Nicolson_1947,LEFORESTIER199159,koonin1990computational} scheme for each Wannier state separately. This implicit method is chosen for its unconditional stability and exact unitarity, ensuring that the total probability $\int |W_n(x,t)|^2 dx$ is conserved throughout the simulation. Note that packet propagation can be carried out independently for each $\Delta E_j$, and only when the calculation of observable $\langle \hat{O}(t)\rangle$ is to be performed, must the data related to all channels  $\Delta E_j$(artificial sub-bands) be merged.  This strategy can be regarded as  effective and stable, as we present in the following.

\section{Towards interacting systems}

Up to now, we have considered the scenario when carriers do not interact. While this assumption  is often justified, the electron-electron (hole-hole) correlations must be included in many numerical modeling tasks. For instance, Coulomb blockade in quantum dots can serve as an archetypical example. Nevertheless, the provided formalism is flexible and can be efficiently adapted to existing many-body numerical methods, such as Matrix Product State (MPS) time propagation or the direct method. We do not focus on the aspects concerning the features of these methods but rather show how to implement packet language into the well known approaches.

Namely, to model the interacting system, we change the paradigm of time evolution. That is, instead of the propagation of Wannier functions, we describe the system in the truncated Hilbert space with basis states represented by $\mathcal{N}$ packets characterized by the quantum numbers $(n,j,\zeta)$, namely emission time, the energy window to which it refers, and chirality. That is, the entire Hilbert space is now  $\mathcal{H}_{\mathcal{N}}=\bigoplus_{n,j,\zeta}\mathcal{H}_{n,j,\zeta}$ and is constructed solely in the wave-packet language with underlying Wannier functions that no longer depend on time but are  \emph{rigid}, and their location and phase depend on $(n,j,\zeta)$. That is why we set $t'=0$ in each $W_{n,j,\zeta}(x,t')\equiv W_{n,j,\zeta}(x)$. Therefore, now, the whole system dynamics is formally encoded in the coefficients $d_{l}(t)$, where the many-body state subjected to time evolution is $|\Phi_0(t)\rangle=\sum_{l}d_l(t)|\phi_l\rangle$. Simply speaking, the assembled Wannier states serve as construction with "slots' whose occupations change according to the time evolution. The states $|\phi_{l}\rangle$ span the Fock space $\mathcal{F}(\bigoplus_{n,j,\zeta} \mathcal{H}_{n,j,\zeta})$; thus, provided $N$ carriers are considered, they are $|\phi_l\rangle=\prod_{i(l)}^{N}\hat{c}_{i(l)}^{\dagger}|0\rangle$ with $i(l)\equiv{(n_i,j_i,\zeta_i)}$. Eventually,  the right ($\zeta=+1$) or left ($\zeta=-1)$ propagating Landauer's current referring to energy window $\Delta E_j$ is restored in the limit $N\rightarrow\infty$ by taking the  expectation value of the initial state  $|\Phi_0\rangle_{N}^{\pm}=\prod_{n}^{N}\hat{c}_{n,j,\pm1}|0\rangle$. Note, however, that the Fock space must be properly equipped with the states related to the $\Delta E_i$ with  $i\neq j$ when scattering centers are present in the system, since the description of  the local excitations cannot be described solely by the $\Delta E_j$ channel in the general case. Eventually, the $\hat{U}(t)=e^{-it\hat{\mathcal{H}}/\hbar}$ is constructed in terms of the desired Hamiltonian, which is 
\begin{equation}
    \hat{\mathcal{H}}=\sum_{ij}(\mathcal{\tau}_{ij}+V_{ij})\hat{c}_i^{\dagger}\hat{c}_j+\sum_{ijkl}\mathcal{V}_{ijkl}\hat{c}^{\dagger}_{i}\hat{c}_{j}^{\dagger}\hat{c}_l\hat{c}_k,
\end{equation}
where the sum of spatial integrals $\mathcal{\tau}_{ij}+V_{ij}$ is given as
\begin{equation}
\mathcal{\tau}_{ij}+V_{ij}=\langle W_i|\hat{\tau}+\hat{V}(x)|W_j\rangle,   
\label{eq:single_integral}
\end{equation}
and  $\hat{\tau}$ is the single particle kinetic energy operator. The interaction amplitudes are 
\begin{equation}
    \mathcal{V}_{ijkl}=\langle W_iW_j|{\hat{v}}|W_{k}W_l\rangle,
\end{equation}
where ${\hat{v}}$ is the carrier-carrier interaction operator. A brief methodological remark regarding the quantum mechanical \emph{picture} is in order here.  As mentioned for the non-interacting case, the formulation is performed entirely in the Heisenberg picture. However, by taking advantage of packet orthogonality, the problem is technically reduced to the Schrödinger picture for the time evolution of a single (non-interacting) Wannier function. On the other hand, the interacting approach described above is both formulated and solved  in the Schrödinger picture, as the single-particle basis is assumed to be time-independent, which naturally results in time-independent field operators.

As one is usually interested in the particular energy range for which the transport is investigated, we divide the entire truncated Hilbert space into at least three energy intervals (sub-bands). That is, the bottom sub-band $\epsilon_b\in(0,E_f)$, the transport sub-band $\epsilon_t\in[E_f,eV)$, and the upper sub-band $\epsilon_u\in[eV,E_{max})$, where $E_{max}$ is chosen arbitrarily. Such a division is both a pragmatic and necessary choice since the shape of the scattering potential $V(x)$ is ultimately encoded   in  $V_{ij}(x)=W^{*}_i(x)V(x)W_j(x)$ terms; that is, $V_{eff}(x)=\sum_{ij}V_{ij}(x)/\sum_{i}|W_{i}(x)|^2$. Therefore, $E_{max}$ should be taken large enough to reasonably re-cover the $V(x)$ profile. Furthermore, the inclusion of energies beyond the transport energy window is generally indispensable, taking into account that scattering processes demand the presence of accessible states responsible for the accurate description of associated excitations. 

Note that for each of the sub-bands, one constructs $M\geq N$ right- and $M$  left going packets, which results in a truncated Hilbert  space $\mathcal{H}_{\mathcal{N}}$ spanned  by $\mathcal{N}=6M$ single particle states, which subsequently serve to construct Fock's space for $N$ considered particles. Indeed, the system under consideration maps in the second quantization to the finite ladder-lattice of dimensions   $(3\times2)\times M$ or $(3\times2\times2)\times M$ for spinless or spinful fermions, respectively. This directly comes from the construction of truncated Hilbert space, where each single particle is distinguishable by the time-injection index $n\in\{0,1,2,\dots,M\}$, band index $\{\epsilon_b,\epsilon_t,\epsilon_u\}$, chirality $\zeta\in\{+1,-1\}$, and optionally the $z$-th component of spin.

Importantly, the entire framework provided previously regarding the calculation of correlation functions is still operative here. Namely, the matrix $P_{\mu\nu}(x)$ can be computed by using the equation 
\begin{equation}
\begin{split}
    P_{\mu\nu}(x)=\frac{1}{N}\int_{x}^{\infty}W_{\mu}^*(x')W_{\nu}(x')dx'\times \\ \sum_{l,k}\langle\phi_l|\hat{c}_{\mu}^{\dagger}\hat{c}_{\nu}|\phi_k\rangle \lim_{t\rightarrow\infty}d_l^{
    *
    }(t)d_{k}(t).
\end{split}
\end{equation}
Particularly, by performing the unitary transformation of $P_{\mu\nu}$  back into the energy domain, that is
\begin{equation}
    \rho(E,E')=\frac{1}{N}\sum_{\mu\nu}e^{-iEt_{\mu}/\hbar}P_{\mu\nu}e^{iE't_{\nu}/\hbar},
\end{equation}
one obtains the transmission coefficient as $T(E)=\rho(E,E)$. 

In the numerical implementation of the simulation procedure, one is not able to consider an \emph{exact} propagation of the Landauer current, since, by construction, the finite basis is always incomplete in this regard. However, the approximation is  controllable by increasing $\mathcal{N}$; moreover, even calculations for finite series of fermions injected can provide crucial information regarding the dynamics of the system. From a technical point of view, the approach described above perfectly matches both direct evolution and MPS based algorithms. While the former is rather limited to a few particles, the latter can handle a substantially larger number of carriers. Moreover, the mapping onto MPS is  natural as it is formulated in the same algebraic language. Also, assuming that carriers, before being scattered by, e.g., a quantum dot, do not interact as they are part of a non-interacting electron gas medium, the interactions become  local. As tensor-network-based methods are tailored to handle  strong local electronic correlations, the packet framework fits this paradigm.

\section{Numerical results}
In the following, we first present the application of the described formalism, starting from the reconstruction of the Landauer DC state. Subsequently, we analyze the text-book examples of transmission through a single Gaussian barrier and resonances in the two-barrier 1D system, where we discuss the formation of Non Equilibrium Steady State (NESS) and confront it with data obtained directly from the Quantum Transmitting Boundary Method (QTBM). In this manner, we not only establish the \emph{benchmark} of the method but also present how the charging/discharging processes evolve in time within a coherent many-body fermionic state. 

Next, we turn to a scenario where direct time-domain treatment is fruitful. That is, we focus on the  Resonance Tunneling Diode (RTD) at finite temperature, subjected to the incidence of time-varying amplitude of current. In this manner, we demonstrate how this coherent-packet-train methodology can be used in real-life research tasks. 

The last illustration concerns the interacting many-body physics in the field of interest. That is, by invoking the truncated rigid (time-independent) Wannier basis formulation described above, we apply the formalism to the  transport through the model of a quantum dot in the regime with the presence of Coulomb blockade explicitly in the time domain.

\subsection{Reconstruction of Landauer DC state in parabolic regime}

Exploring the common parabolic dispersion $E(k) = \hbar^2 k^2 / (2m)$ in view of the reconstruction of Landauer DC state is a natural choice according to its universality in the modeling of transport in mesoscopic systems. In this regime, the general Wannier construction remains consistent with Eq. (\ref{eq:general_wannier}), where the energy-dependent density of states $\nu(E)$ introduces non-linear phase accumulation. 

\begin{figure}[ht]
    \centering
    \includegraphics[width=0.99\linewidth]{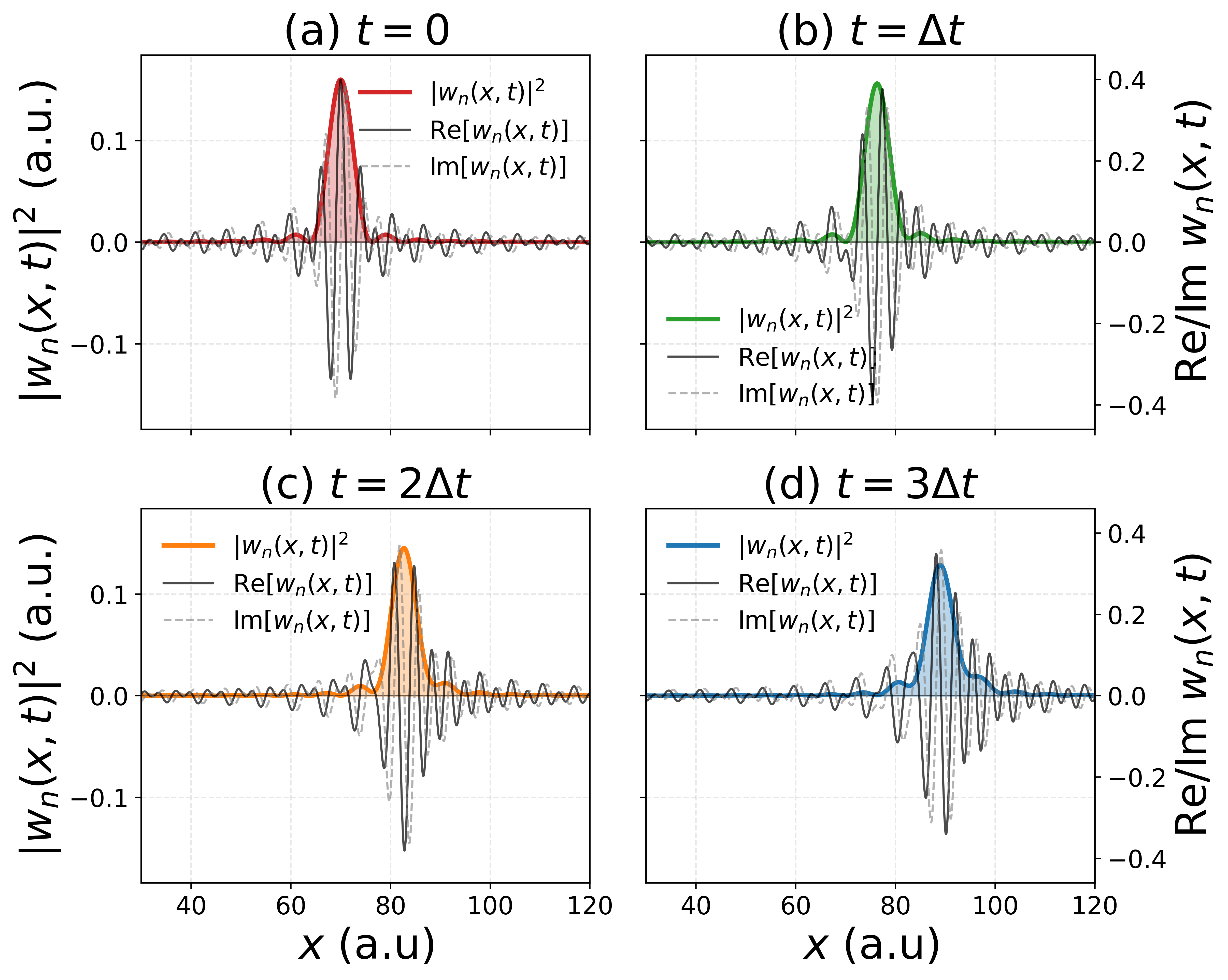}
    \caption{Spatiotemporal evolution of a single Wannier wave packet under a parabolic dispersion relation. Panels (a)--(d) display snapshots of the propagating wave packet at successive time steps $t = 0$, $\Delta t$, $2\Delta t$, and $3\Delta t$. The thick colored curves denote the probability density envelope $|w_n(x,t)|^2$ (left axis), whereas the thin solid black and dashed grey lines represent the real and imaginary parts of the wave function, respectively (right axis).}
    \label{fig:packet_evolution}
\end{figure}

To explicitly capture the wave packet dynamics, we evaluate the Wannier basis numerically. The states at $t=0$ are constructed according to the definition 
\begin{equation}
    W_n(x,0) = \frac{1}{\sqrt{eV}} \int_{0}^{eV} \sqrt{\nu(E)} \, e^{i \left[ 2\pi x \int^E \nu(E') dE' \right]} dE,
\end{equation}
where the integral in the exponent, $k(E) = 2\pi \int^E \nu(E') dE'=\hbar^{-1}\sqrt{2mE}$, accounts for the parabolic mapping between energy and momentum. Here, the factor $\sqrt{\nu(E)}$—which is inversely proportional to the square root of the group velocity $v_g(E)$—ensures the correct spectral weighting of the wave packet components across the injection window.
The resulting dispersive broadening, caused by the non-linear energy-momentum relationship, is illustrated in Fig. \ref{fig:packet_evolution}. As the packet propagates, the envelope exhibits progressive spatial broadening and peak attenuation, a direct manifestation of dispersive propagation inherent to the parabolic energy band. Furthermore, the internal phase oscillations dynamically evolve at a phase velocity distinct from the envelope's group velocity.

\begin{figure}[ht]
    \centering
    \includegraphics[width=0.99\linewidth]{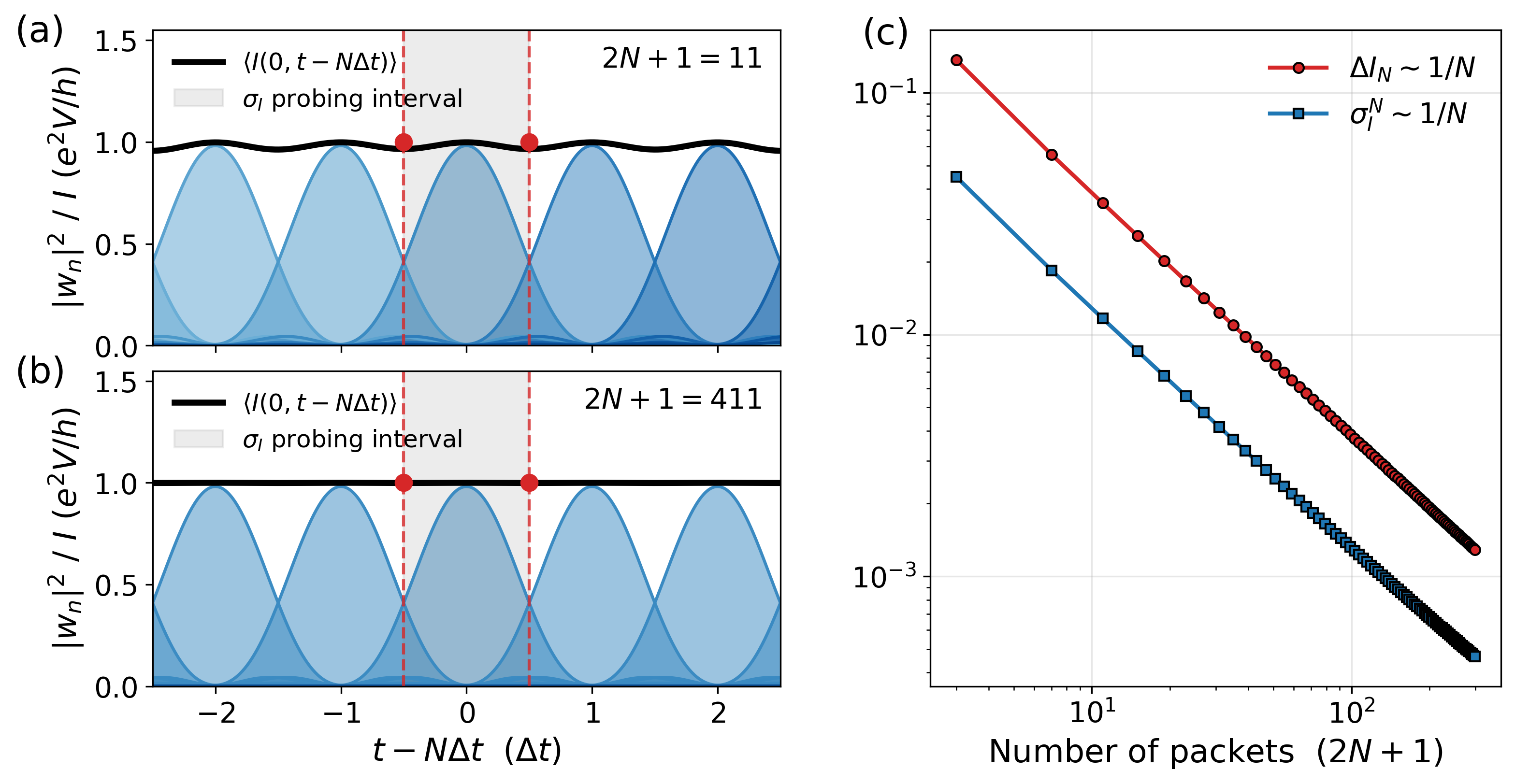}
    \caption{Reconstruction of the direct current (DC) via Wannier wave packet. (a, b) Time evolution of the injected packet probability densities ($|w_n|^2$, blue curves) and the resulting total current $\langle I \rangle$ (solid black lines). Red dashed vertical lines denote the steady-state sampling window. For a sparse packet train ($2N+1=11$, panel a), discrete source oscillations are prominent. Increasing the packet count ($2N+1=411$, panel b) suppresses these fluctuations, smoothly recovering the continuous Landauer DC limit. (c) Log-log convergence analysis. Both the absolute error of the reconstructed current $\Delta I_N$ (red dots) and the standard deviation of the steady-state oscillations $\sigma_I^N$ (blue squares) scale as $\sim 1/N$, confirming the rigorous asymptotic convergence of the proposed time-domain method.}
    \label{fig:k2_DC}
\end{figure}

We have also directly performed the analysis regarding the \emph{quality} (that is, the magnitude of fluctuations) in current by inspecting the role of the number of packets applied in the signal. The conclusive results are presented in Fig.\ref{fig:k2_DC}. By considering $N\gtrsim10^2$, we obtain $\sigma_{I}^N\lesssim10^{-2}$ since it behaves as $\propto1/N$ (see Fig.\ref{fig:k2_DC}c). 

\subsection{Application to the multiple resonant states}

Besides an intuitive, albeit rigorous, real-space-oriented representation of DC formation in a \emph{corpuscular} manner, it also provides the opportunity to perform scattering simulations effectively. Namely, the maximally localized packets are orthogonal and emitted at most with the frequency $f_{max}=1/\Delta t=eV/h$. Assuming that the Hamiltonian is hermitian, the propagator $\hat{U}(t)$ is unitary and preserves the orthogonality of packets. As packet emission events are governed by frequency $f_{max}$, the relation 
\begin{align}
W_{n+m}(x,t)=W_n(x,t-m\Delta t),
\end{align}
holds. Therefore, for this stationary-Hamiltonian scenario, one may perform only a single simulation for the very-first packet $W_0(x,t)$. That is, standard Crank-Nicolson evolution in time and space can be done with high resolution to eliminate numerical artifacts. Then, one is able to compute the single particle observable expectation value $\langle\hat{O}\rangle(t)$ by applying
\begin{align}
\langle\hat{O}\rangle(t)&=\sum_{j=0}^N\int W_j^{*}(x,t)\hat{O}W_j(x,t)dx \nonumber \\
&=\sum_{j=0}^N\int W_0^{*}(x,t-j\Delta t)\hat{O}W_0(x,t-j\Delta t)dx,
\end{align}
where $N$ is the number of packets involved. Eventually, numerical complexity with respect to $N$ for the time-space evolution stage is $\mathcal{O}(1)$ always.

\begin{figure}
    \centering
    \includegraphics[width=0.99\linewidth]{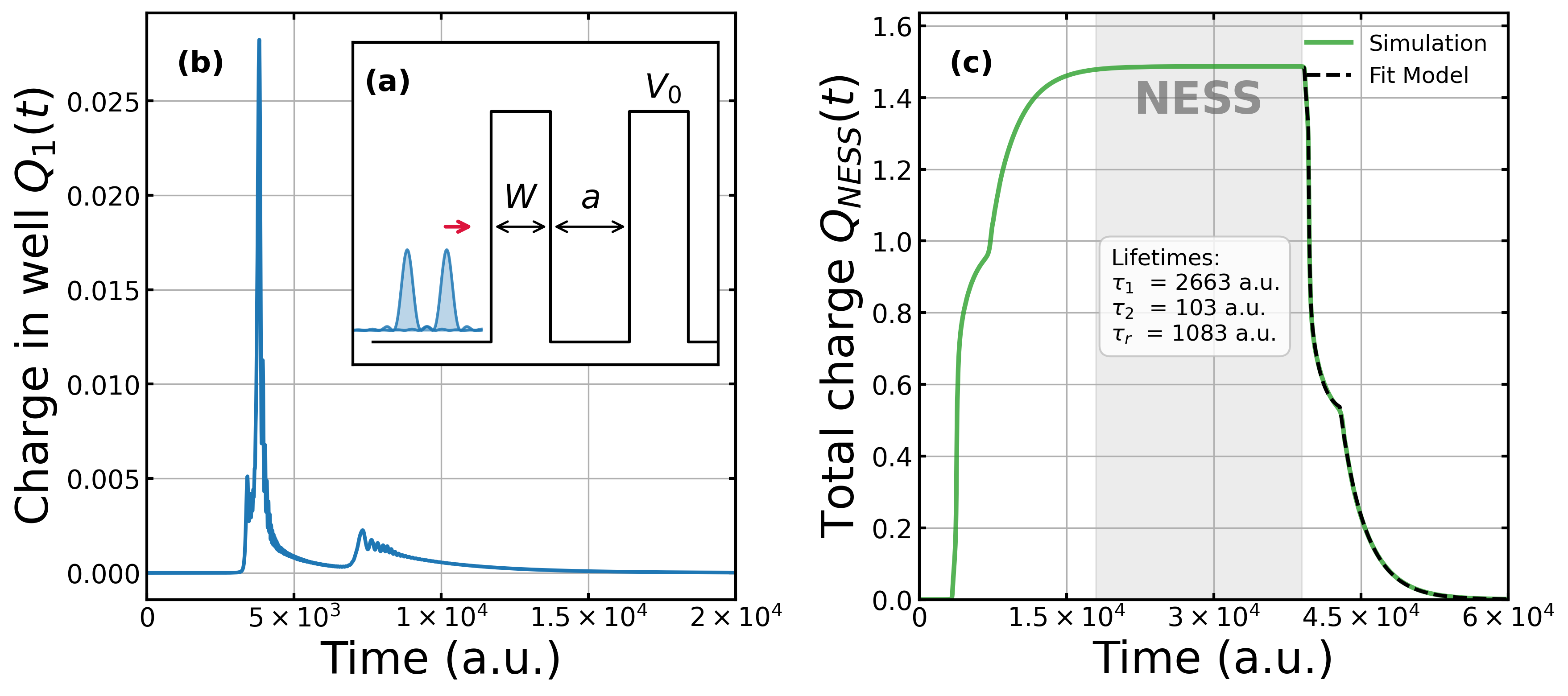}
    \caption{(a) The scheme of the text-book example system subjected to the series of wave-packet (b) Charge $Q_1(t)$ cumulated in the well resulting from the injection of the single packet (c) The formation of NESS during Wannier train scattering. The time decay constants $\tau_i$ are fitted for the range of time which refer to vanishing of $Q_{NESS}(t)$ plateau. }
    \label{fig:experiment}
\end{figure}

Firstly, we utilize this approach to the formation of multiple resonant states in the archetypal 1D system. That is, we study transport through an unperturbed double-barrier structure as shown in Fig.\ref{fig:experiment}a; thus, we assume \emph{a linear response regime} in this case. This is a particularly suitable choice since the direct dynamics of charge density can be studied, and the resultant lifetime for each resonance can be compared to that estimated from a stationary solution by using the Quantum Transmitting Boundary Method (QTBM), which is a standard tool for realizing the Landauer approach in the energy domain. 

In the following, we set $\hbar=m=e=1$; the width of barriers $W=2.5$ and the distance between them $a=4$ are given in arbitrary units, whereas the $eV$ energy window is provided as a fraction of the barrier height $V_0=1.0$. The computational box is of size $L=25\times 10^3$ (a.u.) with $\Delta x=0.25$ (a.u.) spatial resolution. The time $t$ is set in $\Delta t=1/V_0$ units. We also \emph{locally} break the hermiticity near the border of the computational box by adding CAP $iW_{CAP}(x)$ in the standard way, that is 
\begin{equation}
W_{CAP}(x) = \begin{cases}
A_{cap} \cdot \left( \frac{x_L - x}{x_L} \right)^3 & \text{for } x < x_L \\
0 & \text{for } x_L \le x \le x_R \\
A_{cap} \cdot \left( \frac{x - x_R}{L - x_R} \right)^3 & \text{for } x > x_R
\end{cases},
\end{equation}
with $A_{cap}=V_0/5$, $x_L=2.0\times 10^3$, and $x_R=18.0\times 10^3$. 
We intend to inspect the lifetimes of the bound-state resonances (BSR), which are formally \emph{non-equilibrium steady} states (NESS) in the time domain picture. Therefore, we start with the numerical calculations by setting the large $eV=(0.11,0.7)$ to investigate whether all of the BSRs can be characterized in a single simulation. This choice of $eV$ range is reasonable since, in the resulting well, all of the bound states refer to energies $E_1\sim0.17$ and $E_2\sim0.62$. Taking such a large $eV$, we do not physically meet the requirements of the linear response regime; thus, we consider this approach purely as a \emph{numerical experiment} in the hope of observing resonant state decays in a single simulation.

The main results concerning this approach are presented in Fig.\ref{fig:experiment}. Namely, we first focus on the dynamics of the single front-packet emitted at time $T_1=0$ (Fig.\ref{fig:experiment}b). We distinguish two distinctly formed peaks in charge $Q_1(t)\sim\int_{well}dx|W_1(x,t)|^2$ accumulated in the well. The first, appearing $t\sim4\times10^3$, is mainly related to the state referring to $E_2$, since the higher energy \emph{portions} of the packet arrive first at the beginning of the double-barrier system. The height of this peak is naturally greater in magnitude than the preceding one since, excluding energies close to the resonant energies, the probability of transmission generally increases to attain the classical limit for $E>>V_0$. The second peak is clearly affected by periodic fluctuations, which can be viewed as an \emph{echo} coming from the side lobes of the packet. Eventually, the lifetimes of the resonances are hard to compute when relying on single packet dynamics. However, in our simulation, the number of Wannier functions is taken as $4000$—each emitted within the interval $\Delta t\sim 1/eV$. Therefore, they form a non-interacting many-body state conforming to the Pauli exclusion principle, which effectively describes quasi-DC. The term \emph{quasi} stems from the fact that the $I(t) \sim eV/h$ is only fulfilled locally in time and space. That is, the Wannier-train resembles an asymmetric, time-energy dispersion affected kind of \emph{table mountain} due to the finite number of packets taken into account. 

\begin{figure}
    \centering
    \includegraphics[width=0.9\linewidth]{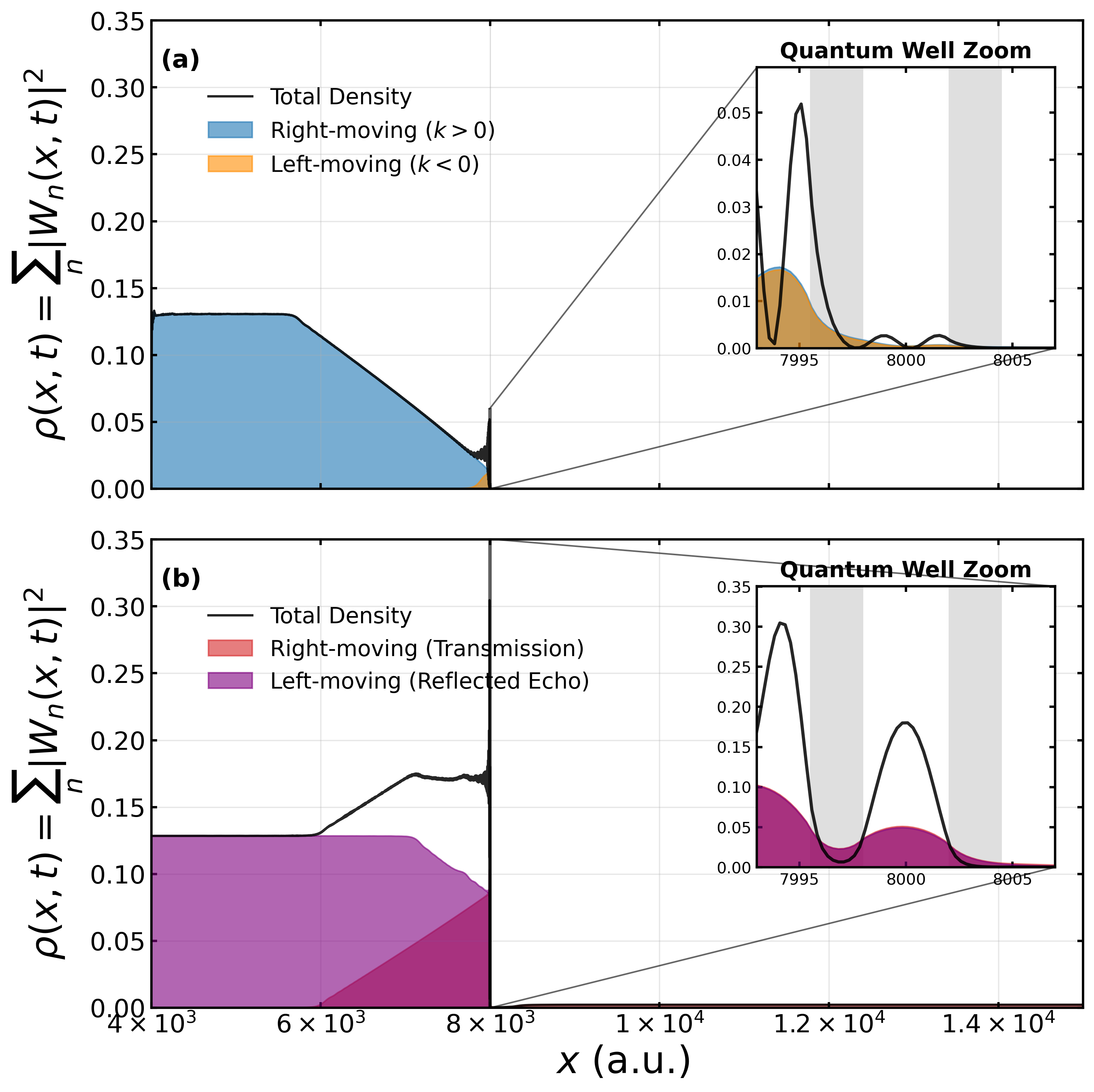}
    \caption{Spatial distribution of density $\rho(x,t)=\sum_n|W_n(x,t)|^2$, computed close to the first-incident time (a) and when decay of NESS starts to proceed (b).
    In the insets we show the zoom onto the two-barrier region to visualize the charge distribution residing inside the well.} 
    \label{fig:tablemountains}
\end{figure}

Nevertheless, the flat-top of this mountain is persistent enough to observe the formation of NESS in the time domain, as shown by the data presented in Fig.\ref{fig:experiment}c, where we plot the total charge in the well as a function of time. In this many-body analysis, one finds that the charging and discharging actions are separated by the NESS plateau -- which extends for $t\in(\sim18\times 10^4, \sim39\times10^4)$ — and appear to consist of at least two distinguishable processes. This explanation perfectly fits the results contained in Fig.\ref{fig:tablemountains}, where the spatial charge distribution is presented at a time closely referring to the first incidents when packets meet the barrier (Fig.\ref{fig:tablemountains}a) and when the steady state (NESS) begins to disappear (Fig.\ref{fig:tablemountains}b). By performing the Fourier transform to the time domain, we also identify the reflected part of the Wannier train by distinguishing $k<0$ and $k>0$ ingredients. As follows from Fig.\ref{fig:tablemountains}a, the resonance related to $E_2$ appears first as claimed before since $|\psi(x\in \text{well})|^2$ forms a $\sim \sin^2(\sim2a)$ profile. 
Eventually, we deconvolute the discharging processes $Q_{t}(t)$ profile into finite Laplace series by taking
\begin{align}
    Q(t)=C_1e^{-t/\tau_1}+C_2e^{-t/\tau_2}+C_re^{-t/\tau_r},
\end{align}
where $\tau_{1,2}$ refers to the lifetimes of resonances, and $\tau_r$ is assumed to be \emph{a residual} ingredient originating from both the decaying slope in \emph{the table mountain} profile and off-resonance transport, which convolutes with resonant-state decay. Eventually, we obtain $\tau_1\approx2663$, $\tau_2\approx103$, and $\tau_r\approx1083$. This identification can be made by inputting auxiliary knowledge regarding bound state energies $E_1$ and $E_2$, or simply by calculating the transmission coefficient $\mathcal{T}(E)$ using the QTBM method or single Gaussian dynamics simulation. However, as we intend to show the complementary character of this method when juxtaposed with the L-B formalism at $T=0$, we follow an alternative approach. As established above, the stationary-Hamiltonian scenario allows the entire many-body transport dynamics to be decoded from the time-evolution of a single Wannier packet, $W_0(x,t)$. Consequently, the energy-resolved transmission coefficient $\mathcal{T}(E)$ can be extracted directly from this single-particle dynamics without invoking stationary scattering states and identifying the resonant energies in the standard manner, that is, by localizing the peaks in $\mathcal{T}(E)$. 

To achieve this, we implement a virtual Time-Flux detector at a fixed spatial coordinate $x_D$ within the free-propagation region of the collector lead ($x_D > x_R$), where the potential $V(x)$ is constant.

\begin{figure}
    \includegraphics[width=0.99\linewidth]{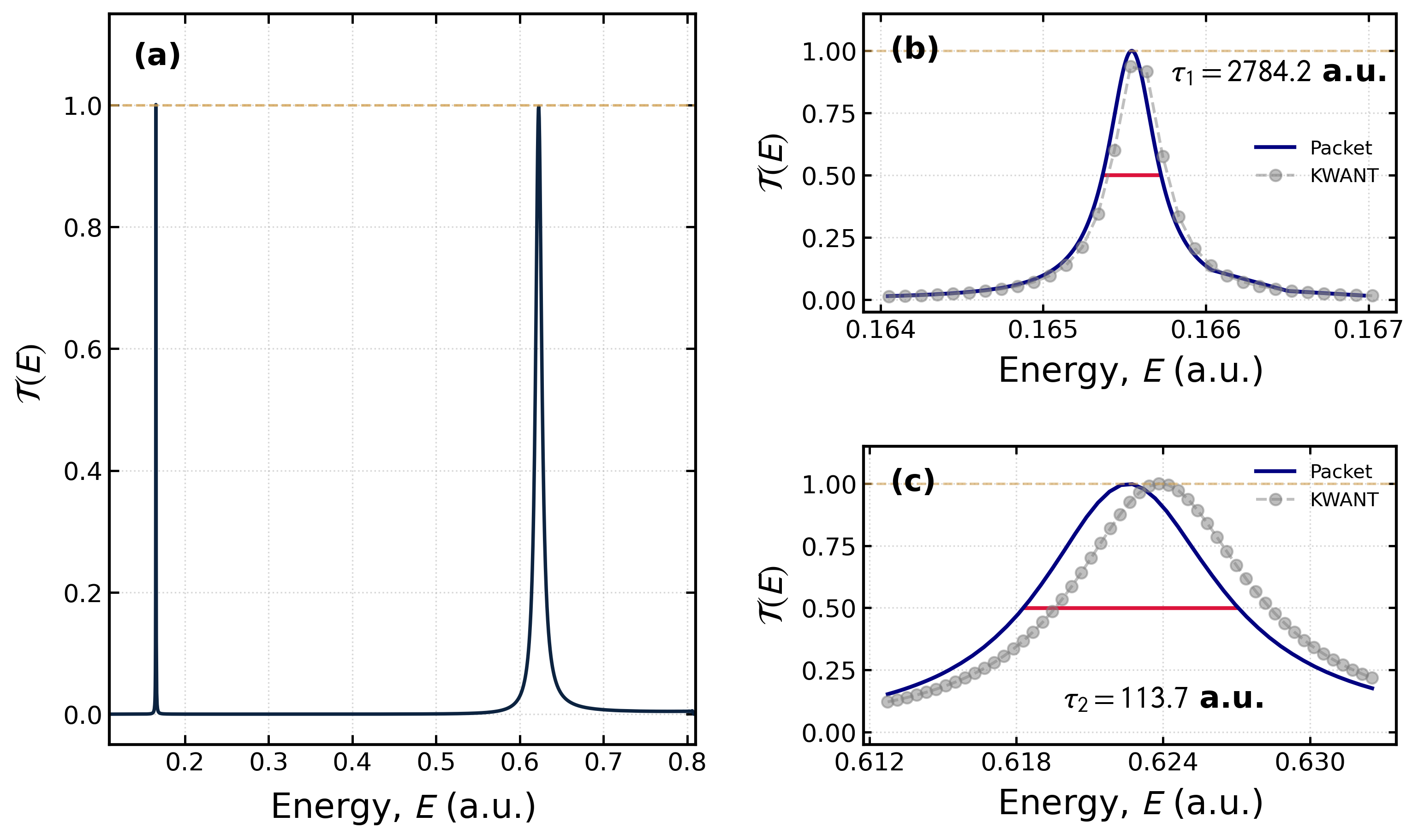}
    \caption{Transmission coefficient $\mathcal{T}(E)$ obtained by means of single packet dynamics. Two distinct peaks are observed for $E_1\approx0.16$ and $E_2\approx 0.62$ (a); By inspecting FWHM for both peaks nearly excellent agreement with fit concerning finite Laplace series in time decays $\tau$'s has been obtained (b,c); For comparison the results obtained by using KWANT package are presented (gray dots in (b) and (c)), fitting exactly to $E_1$ and very closely to $E_2$ (be aware of fine resolution in the insets when compared to the energy window width).}
    \label{fig:tcoeff}
\end{figure}

The outcome of  auxiliary single packet simulation allows us to retrieve the $\mathcal{T}(E)$ profile, as provided in Fig.\ref{fig:tcoeff}. Our findings perfectly match the reasoning given above, since we observe the peaks in $\mathcal{T}(E)$ for $E_1\approx0.16$ and $E_2\approx0.62$. By inspecting the \emph{full width at half maximum} (FWHM) $\Delta E_i^{FWHM}$, for each peak identified, we find $\tau_1\sim1/\Delta E_1^{FWHM}\approx2784$ and $\tau_2\sim1/\Delta E_2^{FWHM}\approx114$
(see Figs.\ref{fig:tcoeff}a,b). Moreover, we have juxtaposed the $\mathcal{T}(E)$ dependency coming from single Wannier dynamics in the vicinity of $E_1$ and $E_2$ with the transmission coefficient resulting from the application of the KWANT package\cite{Groth_2014}. This comparison provides nearly perfect agreement around $E\sim E_1$ and remains accurate for $E_2$, albeit with a minor deviation. This slight discrepancy at higher energies is expected; it arises from the numerical dispersion introduced by the finite mesh resolution, as the kinetic energy operator in our Crank-Nicolson scheme is approximated using finite differences. Despite this tiny deviation, these results successfully benchmark the presented approach against this archetypal system. Importantly, we are able to extract the quasi-bound state decay constants $\tau_i$ directly from the time-dependent simulation, without relying a priori on the energy-time uncertainty relation, since the explicit wave-packet dynamics capture all the essentials of the underlying quantum-mechanical scattering process.

\subsection{Shot noise in transmission through Gaussian barrier}
Yet another illustration of the wave-packet approach in practical implementation is the transport through the Gaussian barrier placed in the ballistic channel. Namely, this simple numerical simulation explicitly shows that any arbitrarily chosen profile of scattering potential can be treated efficiently. Furthermore, it provides the numerical validation of the formula given in Eq. \ref{eq:shot_noise}, which concerns the shot-noise calculation. 

To make these exemplary simulations instructive, we compute  the shot noise $s_{\tau}$ as a function of $E\in(E_f,eV]$. This can be done both by systematically increasing $E$, which leads to the set of energy windows $(E_f,E]$, or one may alternatively divide the entire window into narrow slices of width $\Delta E_n=\delta E$, such that  $eV=\Delta E=\sum_{n}\Delta \delta E $. The second approach is more demanding since slicing implies an enhanced spatial extension of packets. However, it can be tested against such narrow energy windows that the relation $s_{\tau}(E\pm\delta E)\approx\mathcal{T}(E)(1-\mathcal{T}(E))$ should hold, with $\mathcal{T}(E)$ computed within the standard static QTBM approach. Therefore, in the following, we present results obtained with this second approach. We set the scattering potential as $V(x)=V_0e^{-(x-x_0)^2/2\sigma^2}$ with $V_0=0.3$ and $\sigma=2$ (a.u.). As we take $\delta E$ as small as $0.005$, the size of the computational box is large; that is, $L=8\times 10^4$ (a.u.). The interval $\tau$ for which we obtain averaged values of $s_{\tau}$ is selected to refer to the NESS formation stage, meaning that both the increase and decrease in incoming Landauer current stages are excluded. 
\begin{figure}
    \centering
    \includegraphics[width=0.95\linewidth]{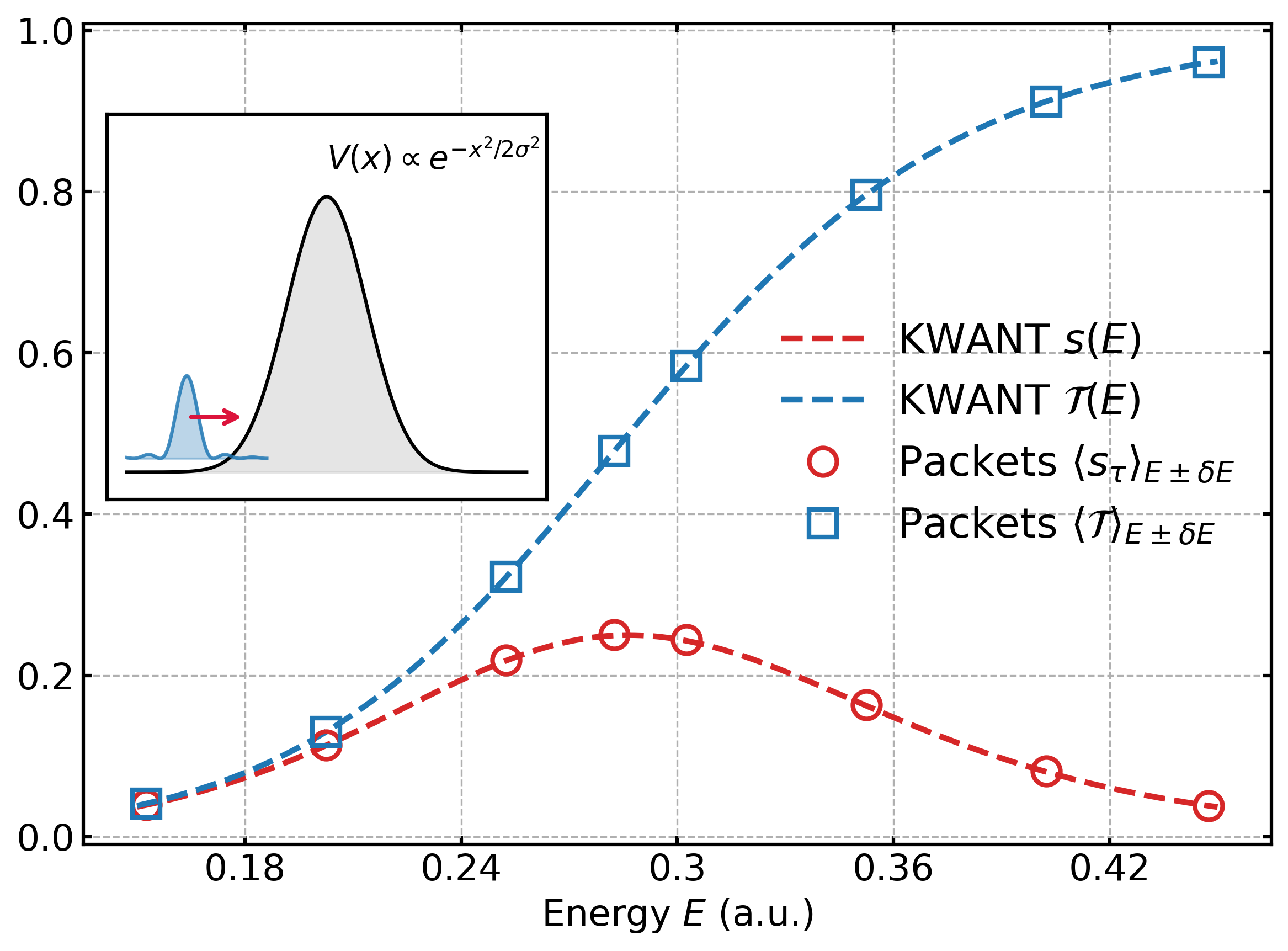}
    \caption{Transmission coefficent and shot-noise at given energy obtained from the simulations performed for the Gaussian barrier as sketched in inset. The data points (symbols) collected from packet approach are confronted again with the solution delivered with the help of KWANT package (lines).}
    \label{fig:shot_noise}
\end{figure}
In Fig.\ref{fig:shot_noise}, we present data obtained from the simulations performed for a Gaussian scattering center, where we plot the transmission coefficient $\langle\mathcal{T}\rangle_{E\pm\delta E/2}$ averaged within each $E\pm\delta E/2$, as well as the shot noise $\langle s_{\tau}\rangle_{E\pm\delta E/2}$. The mentioned averaging directly originates from the construction of Wannier functions within  narrow, albeit finite, energy intervals $E\pm\delta E/2$.   We compare our results with those obtained from the stationary calculations performed with the help of the KWANT package, as previously. We find nearly perfect agreement between both approaches. Namely, the shot noise attains its maximum when $T(E)=0.5$ as  expected. 

This illustrative simulation reveals not only the consistency with standard L-B formalism based on plane-waves but also delivers numerical proof of validity concerning the calculation of  shot noise reformulated into the language of $P_{\mu\nu}$ current matrix elements as defined in Eq. \ref{eq:pmunu}.

\subsection{Resonant Tunneling Diode}

To finalize the examples regarding the non-interacting case, we apply our time-domain NESS formalism to simulate electron transport through a prototypical resonant tunneling diode (RTD) \cite{Sun1998}. This  non-linear device has   been extensively investigated from theoretical perspectives by using a variety of methods \cite{Frensley1987,Frensley1987b,Zang1992,Jauho1994,Ourednik2026}. Therefore, it may serve as a system for which the \emph{proof of concept} of the presented approach can be carried out.  
\subsubsection{Static properties}
\begin{figure}
    \includegraphics[width=0.9\linewidth]{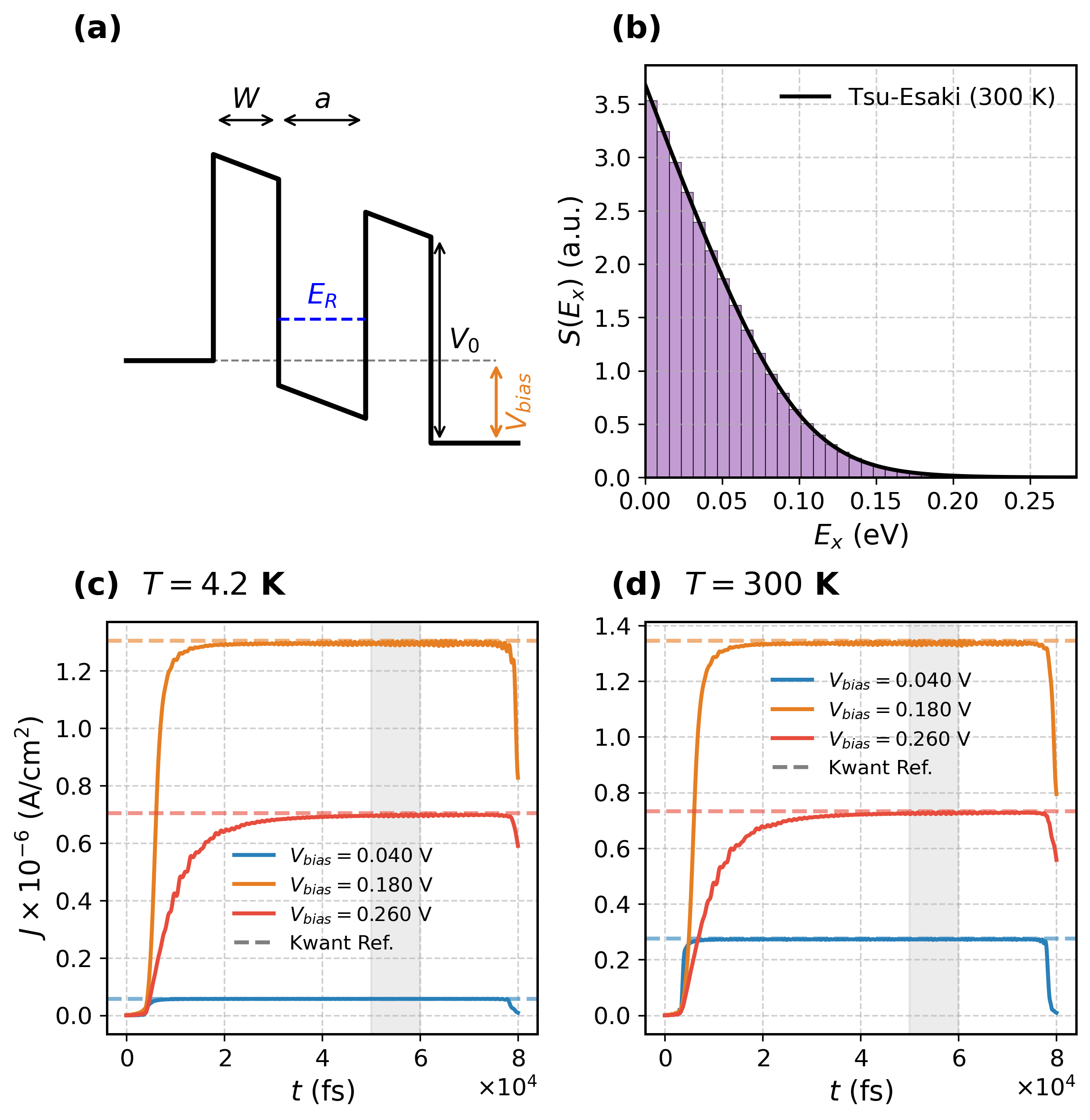}
    \caption{Scheme of RTD diode considered here (a); Slicing in Tsu-Esaki supply function $S(T,E_x)$  at $T_2=300$ K by using $32$ energy windows (bins) which result in parallel simulation of $32$ Wannier trains (b); Current densities $J(t)$ for the three representative biases at $T_1=4.2$K (c) and $T_2=300$K (d). Note, nearly perfect  agreement with data obtained by using code based on KWANT package (dashed lines). The grey-color areas correspond to the time range for which average current has been computed.  }
    \label{fig:rtd1}
\end{figure}

By operating beyond the linear-response regime now, the full three-dimensional system is governed by the separable Hamiltonian:
\begin{equation}
    \hat{\mathcal{H}} = \hat{\mathcal{H}}_0 + \hat{\mathcal{H}}_s + \hat{\mathcal{V}}_{bias} + \hat{\mathcal{H}}_{\mathbf{k}_{\perp}} + \hat{\mathcal{W}}_{cap}.
\end{equation}
Retaining explicit physical units, the longitudinal kinetic operator is defined as $\hat{\mathcal{H}}_0 = -\frac{\hbar^2}{2m^*}\frac{\partial^2}{\partial x^2}$, where we take the effective mass $m^{*}=0.067m_e$, which is a value typical for the GaAs/AlGaAs semiconductors. The scattering region $\hat{\mathcal{H}}_s = \int dx \, V(x) |x\rangle\langle x|$ represents a finite quantum well of height $V_0 = 0.5$~eV formed by two rectangular barriers (width $W = 1.25$~nm, separation $a = 4.5$~nm) as shown in Fig.\ref{fig:rtd1}a. The finite voltage drop between the source and drain is incorporated explicitly via the spatial profile $\hat{\mathcal{V}}_{bias} = \int dx \, V_{bias}(x) |x\rangle\langle x|$, with the overall amplitude denoted as $U_{bias} = e V_{bias}$. As in the previous example, the outgoing wave packets are absorbed by the local complex potential $\hat{\mathcal{W}}_{cap}$. Throughout this analysis, we set the reference source and drain chemical potentials at $T_1=4.2$ K to $\mu_{\mathcal{S}} = \mu_{\mathcal{D}} \equiv\mu = 0.1$ eV. For $T_2=300$ K, we set $\mu_{\mathcal{S}} = \mu_{\mathcal{D}}\equiv\mu= 0.945$ eV to ensure a constant electron density when compared to $T_1$ situation. 

A crucial aspect of our model is the exact treatment of the transverse degrees of freedom. Because the transverse kinetic Hamiltonian $\hat{\mathcal{H}}_{\mathbf{k}_{\perp}} = \frac{\hbar^2}{2m^*}(k_y^2 + k_z^2)$ commutes with the longitudinal part, the full many-body quantum state can be rigorously expressed as a tensor product of the one-dimensional Wannier train propagating along $x$ and the transverse momentum plane waves:
\begin{equation}
    |\Psi\rangle = \left( \prod_n \hat{\mathcal{T}}_n^{\dagger} \right) |0\rangle \otimes |k_y, k_z\rangle.
\end{equation}
This separability allows us to directly extend our time-dependent analysis to finite temperatures $T > 0$. By analytically integrating the 3D Fermi-Dirac distribution over the continuum of transverse modes $|k_y, k_z\rangle$, the statistical weighting of the 1D injected wave packets seamlessly reduces to the well-known Tsu-Esaki supply function $S(T,E_x)\propto\ln{\bigg[1+\exp\bigg(\dfrac{\mu_{\mathcal{S}}(T)-E_x}{k_BT}\bigg)\bigg]}$, where $E_x$ is the kinetic energy associated with transport along the $x$ direction. Consequently, the energy domain integrations correctly span the thermally smeared transport window bounded by the supply function cutoff, mapping our purely 1D time-domain propagation onto the full macroscopic 3D current density.

\begin{figure}
    \includegraphics[width=0.9\linewidth]{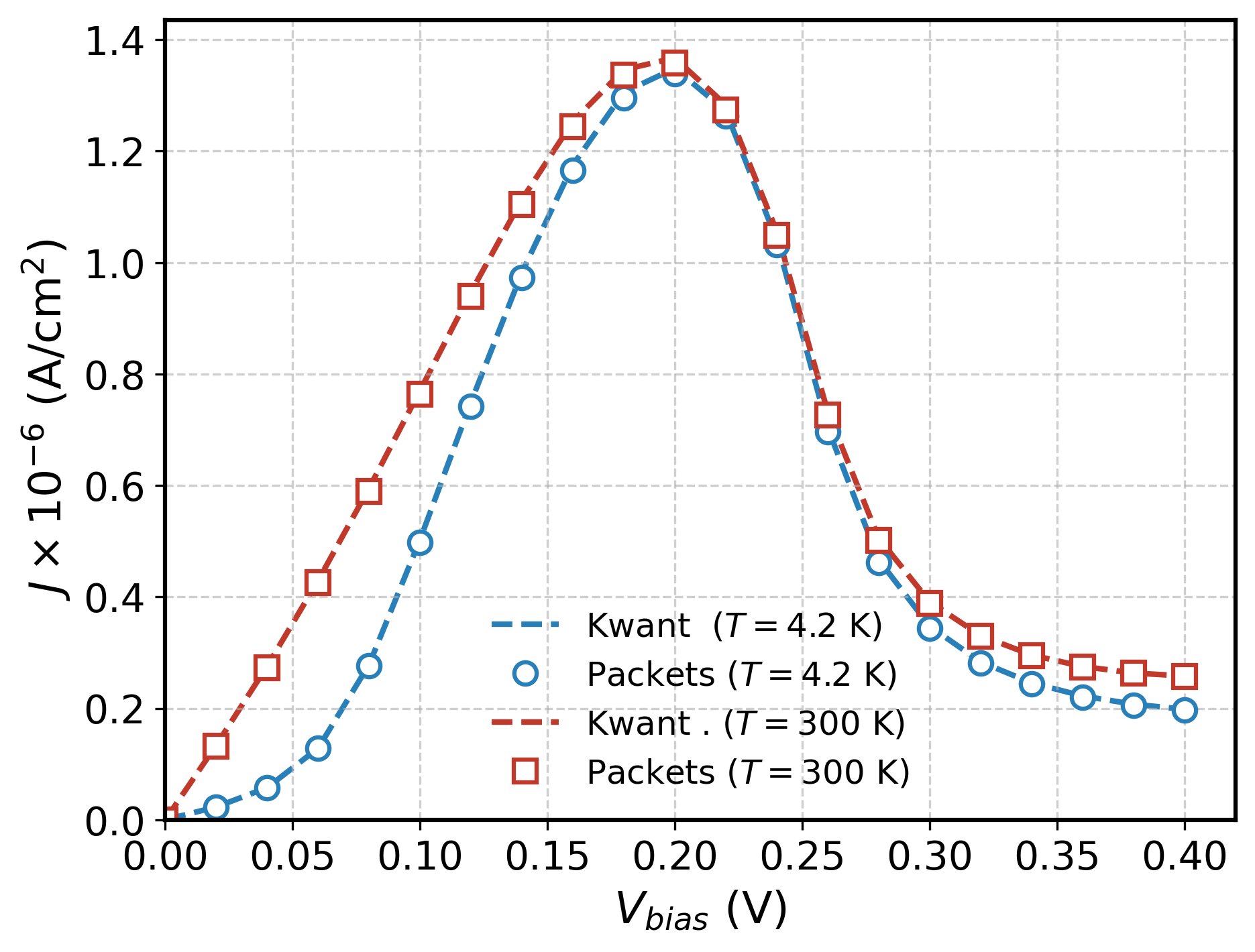}
    \caption{$J(V_{bias})$ dependency for the investigated RTD obtained from wave-packet (NESS) approach (symbols) confronted with static simulations resulting from QTBM as implemented in KWANT package (dashed lines). Current densities are computed as averages over gray areas shown in Figs.\ref{fig:rtd1}c,d.}
    \label{fig:rtd2}
\end{figure}

We start our analysis by confronting the Wannier-train approach to QTBM, relying again on the codes based on the KWANT library prepared specifically for comparison purposes. The archetypal characteristic of the RTD is $I(U_{bias})$ dependency, expressing the famous negative differential resistance. Therefore, as mentioned, we focused on current-bias at two different temperatures, $T_1=4.2$K and $T_2=300$K, to prove the operational capabilities of the orthogonal-packets framework. We divide the transport energy window into  $32$ equal width sub-windows $\Delta E_j$. Such a choice efficiently approximates the shape of the Tsu-Esaki supply function (see Fig.\ref{fig:rtd1}b). The cut-off in the Tsu-Esaki integral is set to $\mu(T)+6k_BT$, which we find  extends sufficiently large for the efficient covering of the thermal tail. In Figs. \ref{fig:rtd1}ab, we present current densities $J(t)$ for the three representative voltage biases at $T_1$ and $T_2$, respectively. One finds that each measured $J(t)$ reproduces the formation of NESS with magnitudes nearly perfectly matching the KWANT package based static simulations. The fluctuations in the signal originate mainly from the trade-off between the number of windows considered and the spatial spreading of Wannier functions when their corresponding widths $\Delta E$ are decreased, which, however, makes the finite temperature simulations possible. Nevertheless, these fluctuations are rather marginal and do not mask the formation of non-equilibrium steady states for both temperatures. Notably, by inspecting the average magnitude of $J$ for the biases $V_{bias}\in\{0.04\text{V},0.18\text{V},0.26\text{V}\}$, we find that, in general, the increase in temperature causes an enhancement in $J$ as expected. Also, $J$ does not behave monotonically with increasing $V_{bias}$, which is just a standard effect in RTDs. Namely, our simulations precisely reconstruct  $J(V_{bias)}$ characteristics by expressing \emph{negative differential resistivity} (NDR) when $V_{bias}\gtrsim0.2$ V, as can be deduced from Fig.\ref{fig:rtd2}. 

\subsubsection{Transport for the time dependent injection rate}

The results presented above can be regarded as the final \emph{proof of concept} for an orthogonal packet based approach. Consequently, having validated our methodology, we turn to the aspect that is beyond the capabilities of static QTBM: the analysis of the RTD response subjected to the time-varying signal.  Namely, in the middle of the simulation run given by $t_M\approx t_{tot}/2$, for the applied $V_{bias}$ and $T$, we switch-on  modulation $\mu_{\mathcal{S}}(t)=\mu+V_{AC}\sin[\omega (t-t_M)]\Theta(t-t_M)$, where $V_{AC}=0.01$eV and $\omega/2\pi=100$ GHz, resulting in  the unidirectional AC propagating from source to drain. The results presented subsequently have been obtained  by using  code that intensively relies on the multi-core architecture and could be adapted to a variety of systems without substantial effort\cite{biborski_zenodo}.
\begin{figure}
    \centering
    \includegraphics[width=0.95\linewidth]{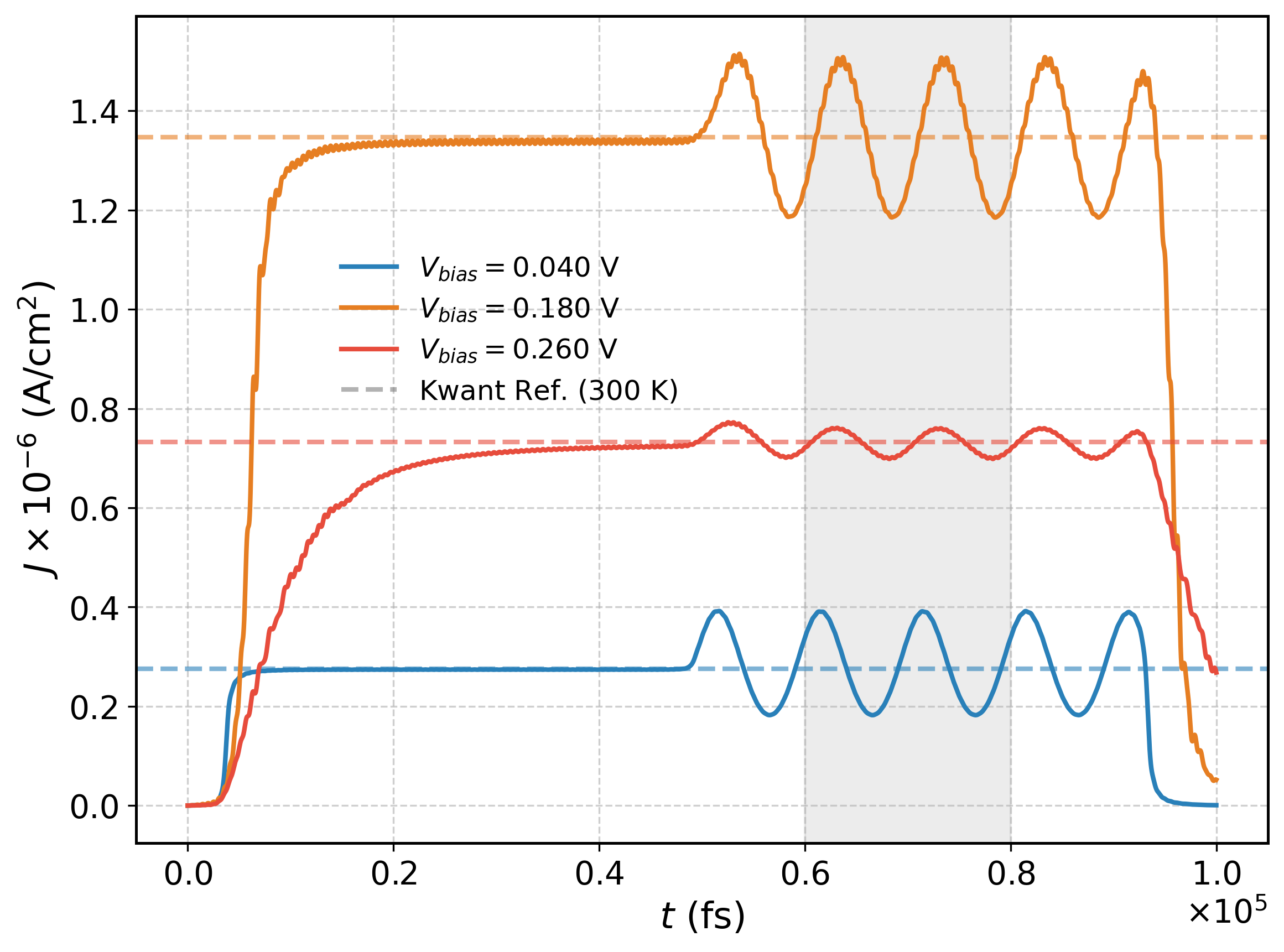}
    \caption{Current densities obtained for the three representative biases at $T=300$K. The varying component in the emitter (source) is switched on $t_M=45000$ fs. Note, that before $v_{ac}(t)$ modulation is applied the NESS corresponds precisely to the current resulting from static QTBM calculations. }
    \label{fig:timevaryingRTD}
\end{figure}

In Fig.\ref{fig:timevaryingRTD} we show the time dependence of current density for the system subjected to the oscillating chemical potential in the source for the three representative biases at $T=300$K. In each simulation, we ensured that a well pronounced NESS emerges before the source oscillation starts. The switch on of sinusoidal amplitude generates the oscillating current of frequency $100$ GHz. The current, especially far from the resonant bias $\simeq0.2$V, exhibits amplitude asymmetry with respect to the base DC, which originates from the higher harmonics in the signal response of RTD. The magnitude of oscillation amplitudes also depends on the bias. Namely, for $V_{bias}=0.04$V, the amplitude is about three times higher than for $V_{bias}=0.26$V, referring to the static NDR regime. This, however, can be explained by deep out-of-resonance transport for the latter  situation. Namely, for $V_{bias}=0.26$V, the resonant state is below the bottom of the conduction band of the source, and the oscillations of $\mu_{\mathcal{S}}$ only slightly influence the total amount of carriers transmitted through the RTD. Contrarily, when $V_{bias}=0.04$ V, the resonant state is in the vicinity of the stationary component of the chemical potential. Most of the carriers participating in the transmission are of energies  close to $\mu_{\mathcal{S}}$. Therefore, each time-driven fluctuation in chemical potential is governed by $v_{AC}(t)= V_{AC}\sin(\omega t)$, influencing the net transmission and, in turn, the resulting current. Naturally, when $V_{bias}\approx0.2$V, the amplitude in $J$ manifests the biggest magnitude since energy-selective time emissions refer to \emph{sweeping} around  resonant energy in the well.

\begin{figure}
    \centering
    \includegraphics[width=0.99\linewidth]{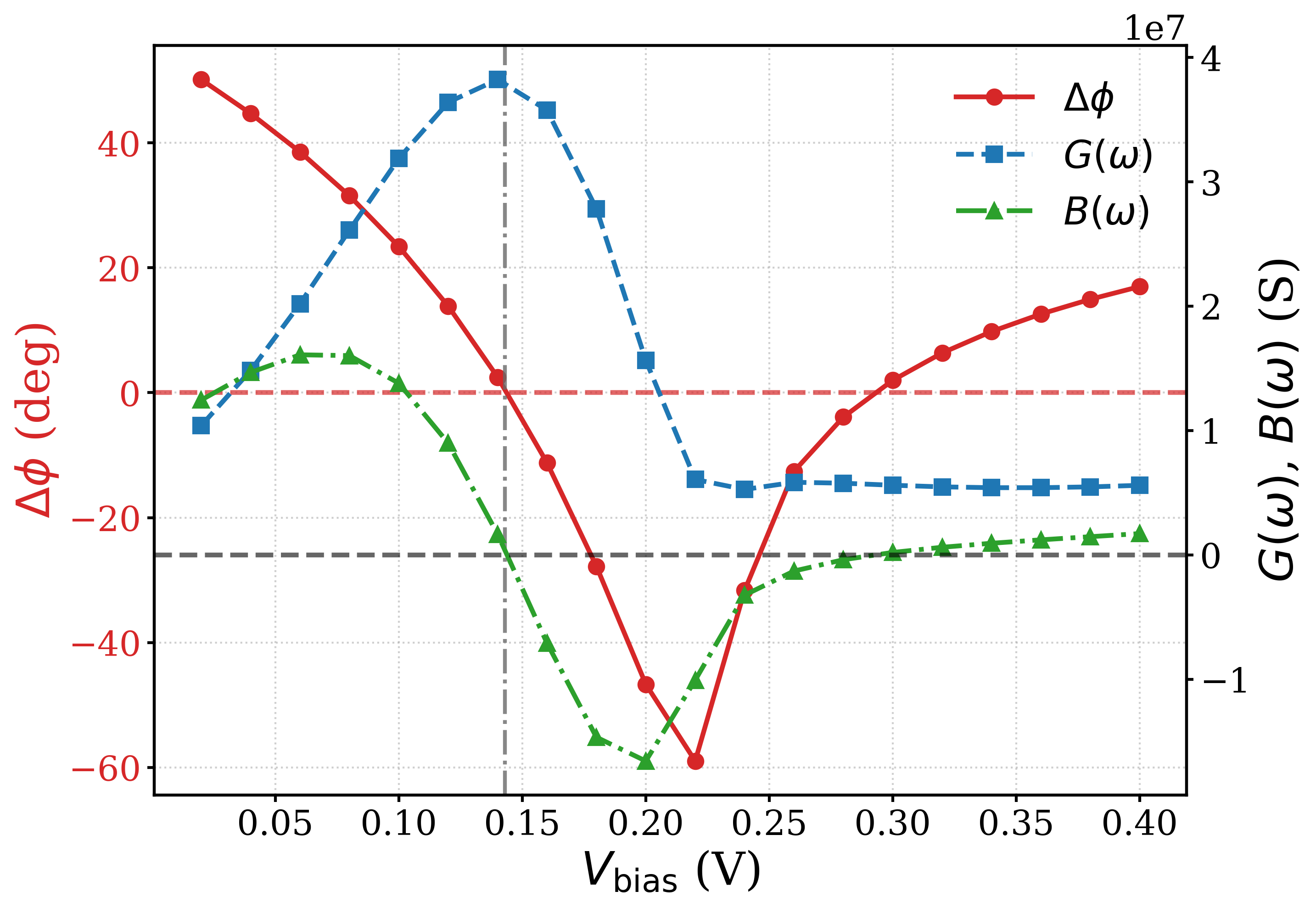}
    \caption{Switch of phase $\Delta\phi(V_{bias})$, $B(\omega)$ and dynamic conductance $G(\omega)$ as a function of $V_{bias}$. Note separate scales for $\Delta\phi$ (left vertical axis) and admittance components (right vertical axis). The dashed line at $V_{bias}\approx0.14$ V indicates change of sign in $\Delta\phi$ and $B(\omega)$ occurring simultaneously.}
    \label{fig:admittance}
\end{figure}

The DC operating point $V_{bias}$ dictates the phase shift $\Delta\phi$ between the driving AC signal $v_{AC}(t)$ and the resulting current $J(t)$. To analyze this dynamical response and extract the complex small-signal admittance, defined in the frequency domain as:
\begin{equation}
    Y(\omega) = G(\omega) + iB(\omega) = |Y(\omega)|e^{i\Delta\phi(\omega)},
    \label{eq:admittance}
\end{equation}
we performed an auxiliary set of calibration simulations, assuming the RTD absence  condition ($V_0 = 0$) in the scattering center. By projecting the resulting time-dependent currents onto the fundamental $1\omega$ Fourier component for both data sets, we evaluated the intrinsic phase shift $\Delta\phi(V_{bias})$. This  procedure rigorously eliminates the trivial propagation delay artifact originating from the spatial distance between the injection point $x_{inj}$ and the RTD structure. With the amplitudes of $v_{AC}(t)$ and $J(t)$ alongside the intrinsic phase $\phi$ isolated, both the dynamic conductance $G(\omega)$ and dynamic susceptance $B(\omega)$ can be determined. In Fig.~\ref{fig:admittance}, we present the extracted intrinsic phase shift $\Delta\phi$, dynamic conductance $G(\omega)$, and dynamic susceptance $B(\omega)$ as functions of the DC operating point $V_{bias}$. The intrinsic phase and the corresponding susceptance exhibit synchronized behavior, signifying a fundamental transition in the dominant tunneling mechanism. 
In the pre-resonance regime ($V_{bias} < 0.14$V), both $\Delta\phi$ and $B(\omega)$ are positive, revealing a capacitive nature originating from the charge accumulation of the incident Wannier wave packets at the emitter barrier. As the source Fermi level aligns with the quasi-bound state of the well, the phase abruptly crosses zero and dives into a deep negative minimum ($\Delta\phi \approx -60^\circ$ at $V_{bias} \approx 0.22$~V). This robust negative phase, macroscopically manifesting as a quantum inductance ($B(\omega) < 0$), is a direct time-domain signature of the quantum dwell time $\tau$ caused by the temporal trapping of electrons inside the resonant structure.

Concurrently, the dynamic conductance $G(\omega)$ remains  positive, acting as a direct probe of the energy-dependent transmission probability $\mathcal{T}(E)$. Notably, $G(\omega)$ reaches its absolute maximum at $V_{bias} \approx 0.14$V, precisely where the susceptance transitions from capacitive to inductive. This peak corresponds to the  point of the steady-state current-voltage characteristics, where the modulation of the source chemical potential $\mu_{\mathcal{S}}$ intersects the steepest slope of the resonance peak, yielding the highest injection sensitivity. 

At higher biases ($V_{bias} \gtrsim 0.26$~V), the resonance level is completely bypassed, and the system enters an off-resonance, direct-tunneling regime. Consequently, $G(\omega)$ flattens out into a stable plateau due to the slowly varying, non-resonant transmission background.  Notably, instead of asymptoting to zero, both the intrinsic phase $\Delta\phi$ and susceptance $B(\omega)$ cross the zero line again and become positive (e.g., $\Delta\phi \approx +17^\circ$ at $V_{bias} = 0.40$V). Since the dynamic conductance $G(\omega)$ remains significantly larger than the susceptance $B(\omega)$, the device does not become a pure capacitor. Instead, the structure essentially operates as a quantum resistor dominated by direct non-resonant tunneling, accompanied by a residual capacitive susceptance originating from the partial accumulation and reflection of incident wave packets at the emitter interface.

\subsection{Interacting packets: Coulomb blockade}
\subsubsection{Model of Quantum-Dot in the interacting regime}
\begin{figure}
    \centering
    \includegraphics[width=0.95\linewidth]{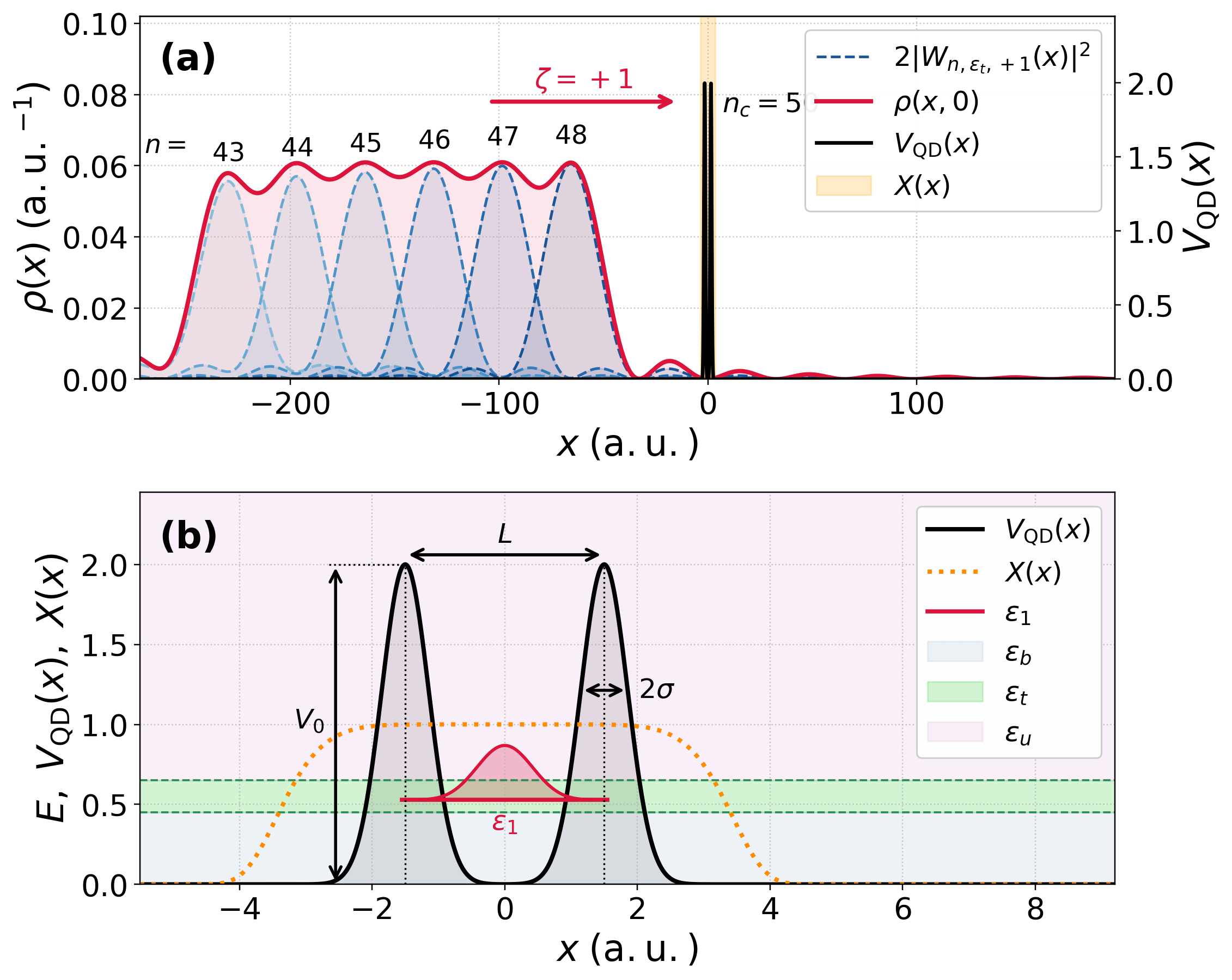}
    \caption{The scheme of the initial stage of transport of finite fermionic train consisting of $2\times6$ orthogonal Wannier packets (a) scattered by quantum-dot which is modeled by the potential given as a linear combination of two Gaussian functions(b). Note, the electron-electron interactions are assumed to be operative only in QD region  as provided by envelope function $X(x)$ (yellow area and yellow dotted line in (a) and (b) respectively) since it is assumed the they are perfectly screened-out in leads.}
    \label{fig:2GAUSS_scheme}
\end{figure}

The last illustration concerns the scenario where electrons experience strong electron-electron interactions in the scattering region. In the following, we consider the system containing two Gaussian barriers, which emulates the cross section of a quantum dot; that is, the potential profile is given as $V_{QD}(x)=V_0\big[e^{-(x-L/2)^2/2\sigma^2}+e^{-(x+L/2)^2/2\sigma^2}\big]$. We set $V_0=2$, $2\sigma^2=1/4$ (a.u.), and $L=3$ (a.u.). In the absence of interaction, this results in two bound (resonant) states with single particle eigenenergies $\epsilon_0\approx0.53$ and $\epsilon_1\approx1.89$. Subsequently, we provide the screened (Yukawa-like) electron-electron interaction potential $V_{e-e}(x,x')= U_0e^{-\kappa|x-x'| }/(|x-x'|+d)$. As carriers in the channel are assumed to be perfectly screened by the immersing medium (electron-gas), we use an effective interacting potential $V_{e-e}^{eff}(x,x')=X(x)X(x')V_{e-e}(x,x')$ where the envelope function $X(x) = \exp\!\left\{-\left[\frac{x}{\frac{1}{2}L + 4\sqrt{2}\sigma}\right]^8\right\}$ smoothly screens out the interactions outside the modeled quantum dot. The sketch representing the aim of the simulation, as well as the details concerning the scattering center, is provided in Fig.\ref{fig:2GAUSS_scheme}.

The exact diagonalization (ED) in the single particle basis performed for this system delivers $E(N=1)=\epsilon_0$ and $E(N=2)\approx1.44$. Since $\epsilon_1>E(N=2)>2\epsilon_0$  we focus on carrier energies $\sim E(N=2)/2$ to investigate the Coulomb blockade associated with the transport solely in the vicinity of $\epsilon_0$. We take into account all  $\mathcal{V}_{ijkl}$ terms in the many-body Hamiltonian, albeit limited to the region of the quantum dot, and density-density interactions in its vicinity (see Appendix \ref{sec:appB} for details), consistent with the shape of the envelope function $X(x)$.
In the following, we set  $d=2$ (a.u.), $\kappa=0.1$ (1/a.u.), and use $U_0$ as a parameter that can gradually switch the system between non-interacting and strongly interacting regimes. Wannier functions in each sub-band have been obtained numerically by taking $4\times 10^4$ points in energy space for each channel and tabulated for $x\in(-2500,2500)$  (a.u.), assuming a spatial resolution of $\Delta x=0.1$. Subsequently, we have used these  fine resolution packets and, through numerical integration, computed all $\tau_{ij}$ and relevant $\mathcal{V}_{ijkl}$ microscopic parameters.

\subsubsection{Application to the time evolution of MPS}
Our \emph{rigid-Wanniers} approach is, by construction, compatible with the state-of-the-art
MPS-based many-body description used extensively within the Density Matrix
Renormalization Group (DMRG) framework in strongly correlated many-body
physics. MPS-based methods have already been successfully applied to modeling
quantum transport in nanodevices~\cite{Chung2022,Schoenauer2019}. Here, we
employ the time-dependent variational principle (TDVP)
algorithm~\cite{Haegeman2011,Haegeman2016} as implemented in the ITensor
library~\cite{SciPostPhysCodeb.4}. We consider the real-time transport of a
finite Landauer spinful pulse defined as a Wannier train consisting of
$N = N_{\uparrow} + N_{\downarrow}$ fermions with
$N_{\uparrow} = N_{\downarrow} \in \{2, 4, 6\}$ ($N \in \{4, 8, 12\}$). In the
presented simulations, we retain $M = 100$ Wannier positions for each chiral
branch of the conduction (transmission) channel $\epsilon_t$. Outside the
scattering region, the asymptotic contribution of the initially unoccupied
bottom and upper virtual bands ($\epsilon_b$ and $\epsilon_u$) to the
transmitted and reflected charges is negligible. Physically, this follows from
two facts: first, because the resonant energy $\epsilon_r \in \epsilon_t$ lies
 outside the energy spectra of bands $\epsilon_b$ and $\epsilon_u$,
single-particle virtual excitations at the barriers form an evanescent dressing
cloud that decays exponentially with the lattice distance $|n - n_c|$ from the
quantum dot; second, the suppression of double occupancy in the quantum dot
driven by Coulomb repulsion simultaneously inhibits inelastic two-particle
scattering processes that could otherwise populate states outside the incident
energy window asymptotically. This observation allows us to substantially
reduce the MPS length by retaining Wannier states from the lower ($\epsilon_b$)
and upper ($\epsilon_u$) sub-bands only in the spatial vicinity of the
scattering center. Mapping the center of the QD ($x_c^{\mathrm{QD}} = 0$) to
$n_c = M/2 = 50$, we include the $\epsilon_b$ and $\epsilon_u$ modes within the
local window $|n - n_c| \leq K_{\mathrm{virt}} = 10$ ($n \in [40, 60]$).
Consequently, the active single-particle basis consists of
$2 \times [100 + 2 \times (2K_{\mathrm{virt}} + 1)] = 2 \times 142 = 284$
spatial chiral orbitals, corresponding to $\mathcal{N} = 2 \times 284 = 568$
spinful Wannier orbitals instead of $2 \times 6 \times 100 = 1200$. 

We find that the stage in which one tailors the MPS ansatz to a given problem
is the most demanding task in the whole procedure. As the validity of the
problem-specific MPS topology is crucial for obtaining a physical outcome, it
should always be benchmarked and treated with care. A reasonable protocol
is to test the elaborated MPS against the non-interacting case  in
the very same single-particle basis. This can be achieved by a direct
construction of the single-particle time-evolution matrix
$U_{ij}(t) = \langle 0 | \hat{c}_{i}\hat{U}(t)\hat{c}_{j}^{\dagger} | 0 \rangle$
and its action on the initially occupied single-particle states.
Accordingly, we benchmarked the approach by taking
$U_0 = 0$. 
We found that partitioning the system into three
functional regions along the  MPS architecture is highly effective. Namely,
outside the quantum dot center we retain the chiral propagating Wannier
representation $W_{n,j,\zeta}(x)$ ($j \in \{\epsilon_b, \epsilon_t, \epsilon_u\}$,
$\zeta \in \{+1, -1\}$); whereas  at the quantum dot center
$n_c = M/2 = 50$ ($x_c^{n_c} = 0$) we introduce a local unitary rotation into
states of definite spatial parity. Formally, the $284$ four-level electronic
MPS sites ($\mathcal{N} = 568$ spinful orbitals) are arranged along the 1D MPS
chain $\mathcal{C} = \mathcal{R}_{\text{in}} \oplus \mathcal{R}_{\text{QD}} \oplus \mathcal{R}_{\text{out}}$
as detailed in Appendix~\ref{sec:appB}.

The real-time evolution of $|\Phi(t)\rangle$ has been subsequently carried out within
a hybrid TDVP scheme that alternates five single-site ($1$-TDVP) time steps
with one two-site ($2$-TDVP) bond-expansion step ($\Delta t = 0.125\text{ a.u.}$,
maximum bond dimension $\chi_{\max} = 1024$, and singular-value truncation
cutoff $10^{-10}$), while explicitly conserving the total particle number
$N \in \{4, 8, 12\}$ and total spin projection $S^z_{\mathrm{tot}} = 0$.
The results presented below can be reproduced by running the codes provided by us \cite{biborski_zenodo2}.

\subsubsection{Numerical results}

\begin{figure}
    \centering
    \includegraphics[width=0.99\linewidth]{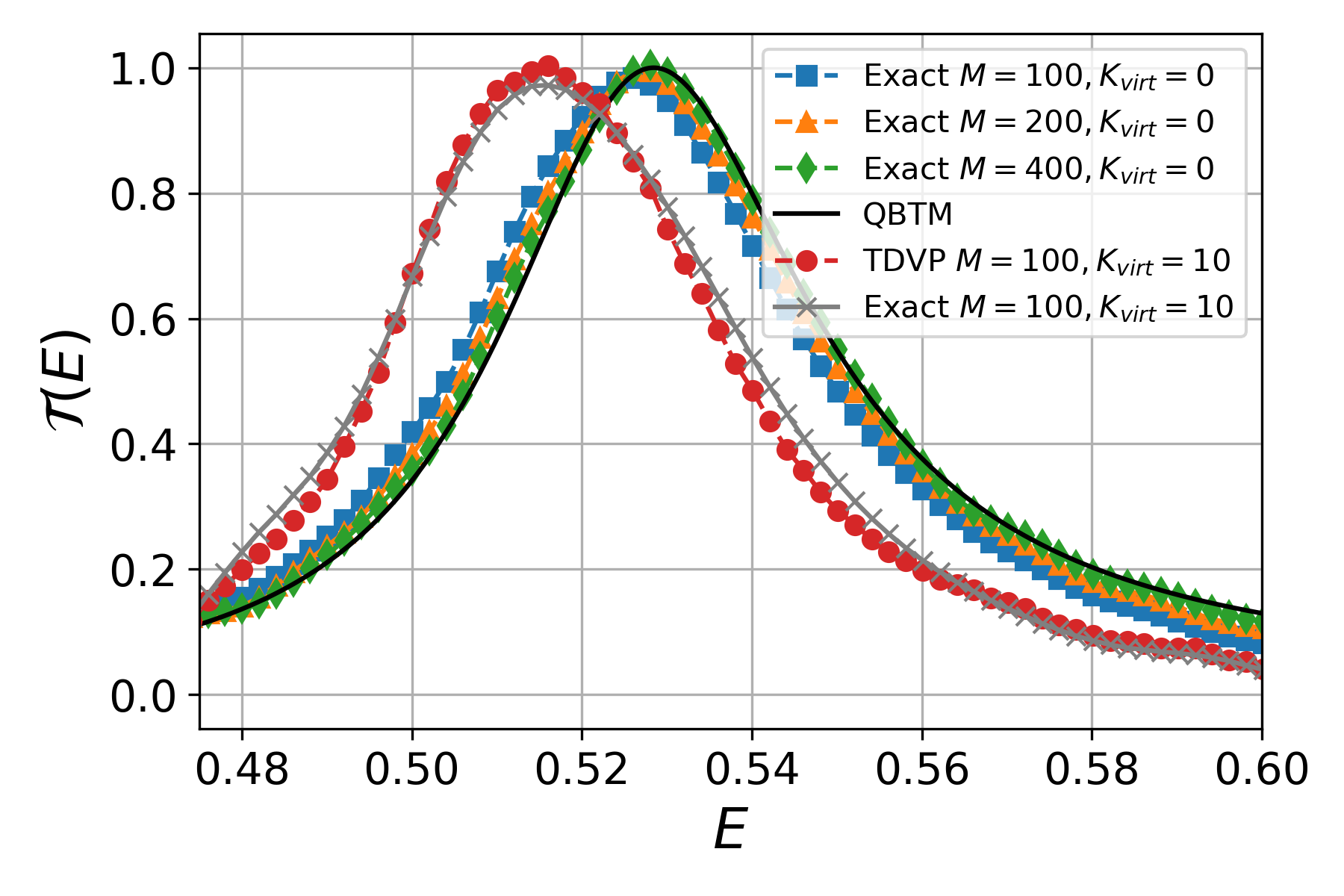}
    \caption{Transmission coefficient $\mathcal{T}(E)$ obtained via stationary single-particle approach (QTBM) and within direct time evolution using exact and TDVP-MPS approaches for $U_0=0$ (non-interacting fermions). The time evolutions (RW-approach) from whose $\mathcal{T}(E)$ curves result,  have been performed for $N=4$ fermions in each case.}
    \label{fig:basis_analysis}
\end{figure}

First, we present the analysis that proves the equivalence of the rigid-Wannier (RW) approach with the standard L-B formalism  and  the impact of basis truncation. For this purpose, we have analyzed transmissions at $U_0=0$ by using QTBM as implemented in KWANT, as well as \emph{exact} and TDVP-MPS evolutions carried out in the RW framework. In Fig. \ref{fig:basis_analysis}, we present the outcome of this analysis. Namely, we find excellent agreement between the exact treatment at $M=400$ and the L-B QTBM based approach operating directly in the energy domain. Subsequent truncation of the basis by the reduction solely of $M$ changes the outcome marginally; that is, the resonance shifts marginally to smaller values of energy, even for $M=100$.  The reduction of basis size, also by the elimination of those Wannier states constructed for $\epsilon_b$ and $\epsilon_u$ sub-bands, which are outside the scattering region ($K_{virt}=10$), results in a more pronounced shift of resonant energy. Nevertheless, for $M=100$ and $K_{virt}=10$, the agreement with the exact solution is still acceptable since the peak  is shifted by $\delta E\lesssim0.01$. Those values of $M$ and $K_{virt}$ correspond to MPS sizes that are tractable for the TDVP protocol in a reasonable time. We find nearly perfect agreement in this truncation scheme with the exact solution, as shown in Fig. \ref{fig:basis_analysis}, which proves the accuracy of the MPS structure. Therefore, as a compromise between physical accuracy and computational effort, we turned to the interacting mode $U_0>0$ for such a scheme of truncation.

\begin{figure}
    \centering
    \includegraphics[width=0.95\linewidth]{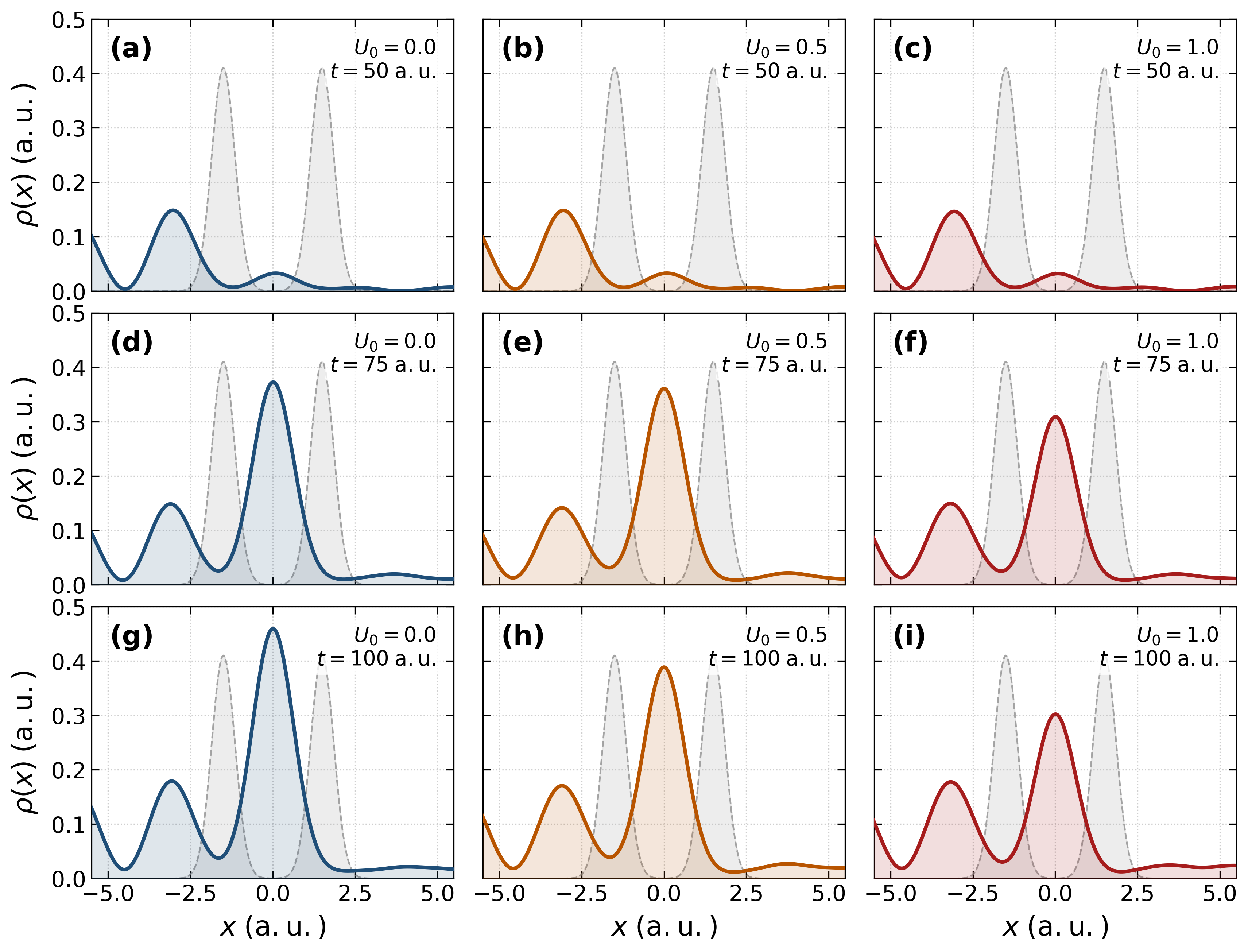}
    \caption{The panel representing the snapshots of electronic density $\rho(x,t)$ within QD range for the three different $t=50,75$ and $100$ (a.u.) (rows) and three different values of $U_0=0.0,0.5$ and $1.0$ (columns) when $N=4$. The reduction of accumulated charge  inside the dot is especially pronounced for $t=100$ (a.u.), at which resonant state develops, since electron-electron interactions becomes mostly functional then.}
    \label{fig:tdvp_panel}
\end{figure}

In Fig.\ref{fig:tdvp_panel} we present snapshots of the spatial distribution of electron density computed as $\rho(x,t)=\sum_{ij}W_{i}^{*}(x)W_{j}(x)\langle\Phi_0(t)|\hat{c}_i^{\dagger}\hat{c}_j|\Phi_0(t)\rangle$ (where $i,j$ implicitly contains information regarding $n$, $\zeta$, and $\sigma$ quantum numbers) for three different values of $U_0=0.0,0.5$ and $1.0$, and three different values of time $t$ during the simulation for $N=4$. We focus on the QD area as the most interesting in view of the emergence of the Coulomb blockade. Notably, one observes that at $t=50$ (a.u.), all three interaction regimes provide a similar distribution of charge since the amount of charge inside the dot is small and electron-electron interactions are not operative (Figs.\ref{fig:tdvp_panel}a-c). Nevertheless, when the charging of an initially empty QD proceeds over time ($t=75$ (a.u.) in Figs.\ref{fig:tdvp_panel}d-f), the cumulation of charge decreases with the increasing value of $U_0$. Eventually, at maximal fillings in the QD, which occur at $t\approx100$ (a.u.), the suppression of charging is clearly visible; that is, the Coulomb blockade becomes operative with increasing $U_0$ (Figs.\ref{fig:tdvp_panel}g-i).

\begin{figure}
    \centering
    \includegraphics[width=0.95\linewidth]{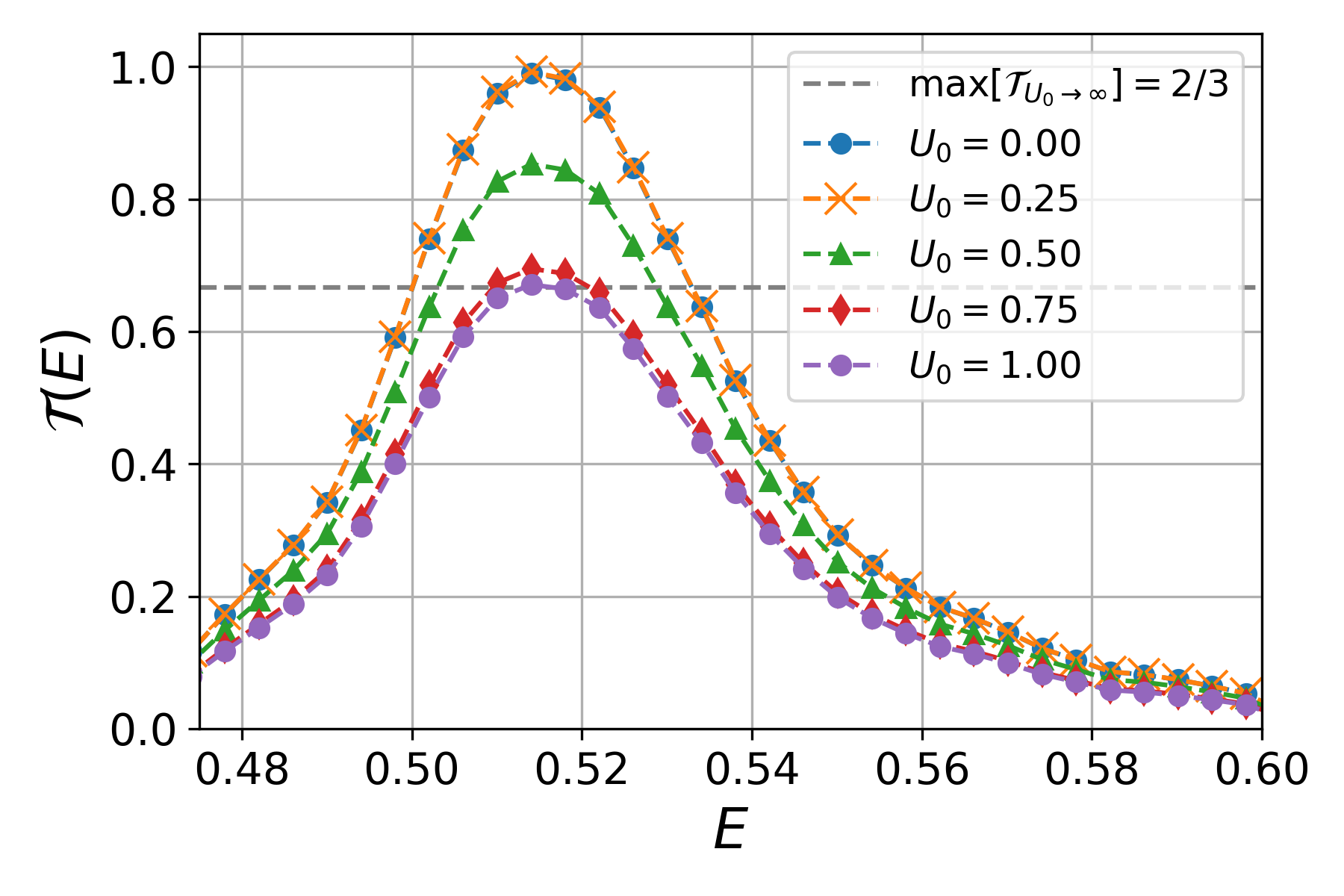}
    \caption{The transmission coefficient $\mathcal{T}(E)$ obtained in TDVP runs for the train consisting of six pairs of spin-up/spin-down packets ($N=12$) at different values of $U_0$. The dashed, gray horizontal line indicates the limit $2/3$, that is, the blocking of transport - driven by electron-electron interactions in QD -  is maximal.}
    \label{fig:T_E_12el}
\end{figure}

\begin{figure}
    \centering
    \includegraphics[width=0.95\linewidth]{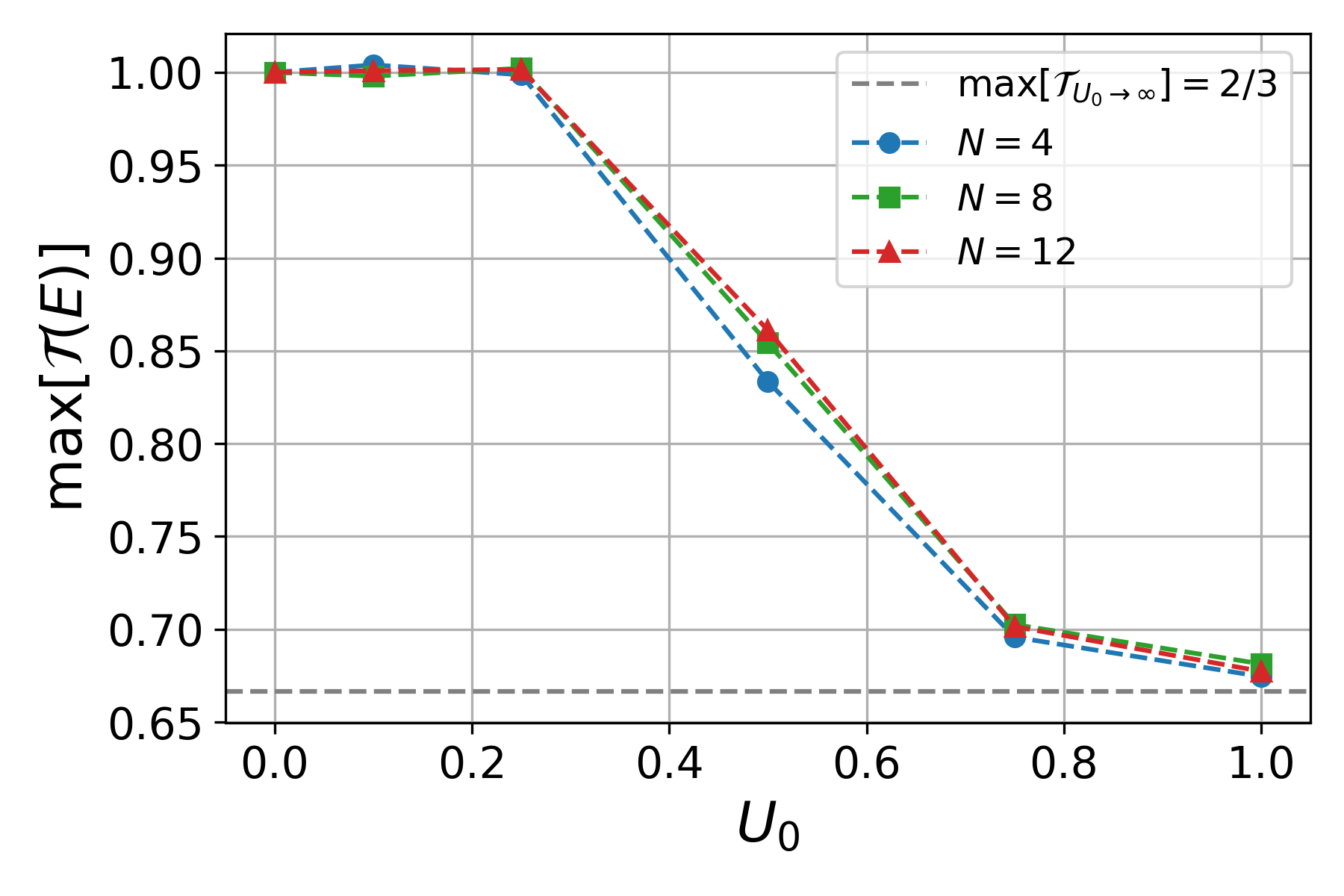}
    \caption{The maximal value of transmission coefficient as a function of interaction magnitude factor $U_0$ for three different sizes of Wannier train.}
    \label{fig:Tmax}
\end{figure}

To quantify the role of the increasing magnitude of interactions for the considered transport window $\Delta E$  with respect to the development of the Coulomb blockade, we subsequently investigated the transmission coefficient  curves for $U_0\in\{0.0,0.25,0.5,0.75,1.0\}$. As considered by Glazman and Matveev \cite{GlazmanMatveev1988}, based on the impurity Anderson model \cite{Meir1991} resulting in a Meir-Wingreen\cite{Meir} type of formula for $\mathcal{T}(E)$, one finds that for the maximally operational Coulomb blockade $\text{max}[\mathcal{T}_{U_0\rightarrow\infty}(E)]=\frac{2}{3}\mathcal{T}_{U_0=0}(\epsilon_r)$,  that is, in the regime  in which the double occupancies in QD are fully projected out by electron-electron interactions. In Fig.\ref{fig:T_E_12el} we show the $\mathcal{T}(E)$ dependencies resulting from TDVP simulations performed for $N=12$ electrons for the selected values of $U_0$. The role of two-body interactions, at least up to $U_0\lesssim0.25$, is marginal; however, at $U_0=0.5$, one observes a substantial decrease in transmission amplitude, although at maximum it is still notably higher than $2/3$. Further increases in the magnitude of $U_0$, however, indicate the formation of a full blockade. Namely, for $U_0=0.75$, the maximum of $\mathcal{T}(E)$ is only slightly higher than the limit predicted by Glazman and Matveev for the continuous stream of carriers with fully projected double occupancies. Eventually, at $U_0=1.0$, we observe the full operational capability of the Coulomb blockade, since $\text{max}[\mathcal{T}(E)]\approx\frac{2}{3}$. We find this behavior to be almost independent of $N$ , as we present in Fig.\ref{fig:Tmax}, where we plot the maximal transmission as a function of $U_0$ for all considered $N$. 

Note that the value of $U_0$ for which the full-switch-on of the Coulomb blockade appears is not independent of the width of the transmission band $\Delta E$. Indeed, we are able to study the mechanism of its emergence by invoking Hubbard band splitting. Let us consider the semi-classically estimated position of the upper Hubbard band \cite{Meir1991}, $E_{UHB}(U_0)\approx\epsilon_0+U_{eff}(U_0)$, where $U_{eff}(U_0)$ can be estimated by calculating the Hartree interaction kernel for QD at 
$U_0=0$ and $t_{c}$, referring to the NESS formation (temporarily maximally charged QD), that is:
\begin{equation}
    U_{eff}(U_0)\approx U_0\int_{x=-2.5}^{x=2.5}\rho_{\uparrow}(x',t_c)V_{e-e}(x',x'')\rho_{\downarrow}(x'',t_c)dx'dx''.
\end{equation}
This estimation  left us with $U_{eff}\approx0.072\times U_0$. The full width at half maximum (FWHM) of the single-particle resonance peak is $\Gamma \approx 0.044$ for the parameters of QD used by us. Therefore, $E_{UHB}\gtrsim\Gamma/2+\epsilon_0$ when  $U_0\gtrsim0.28-0.33$. This accurately explains why we observe an abrupt decrease in $\text{max}[\mathcal{T}(E_0)]$ when $U_0\gtrsim0.25$ (see Fig. \ref{fig:Tmax}). Notably, for $U_0=1.0$, we are close to the upper edge of the $\epsilon_t$ sub-band, for which transmission is substantially reduced when analyzing the non-interacting case. Thus, the simultaneous transport of two carriers is almost totally blocked.

Recapitulating, the numerical evidence for microscopically modeled transport through QD in the interacting regime provided above matches the common understanding of the development of the Coulomb blockade based on the effective (Anderson) model. Therefore, we find the RW approach  applied to the TDVP-MPS numerical scheme to be positively validated in this regard as well.

\section{Discussion and Conclusions}

In summary, the presented wave-packet formalism bridges the gap between the stationary Landauer--Büttiker framework and the explicit time-domain dynamics of fermionic excitations. While standard localized approaches \cite{FEIT1982412,DION2014407}(e.g., utilizing single Gaussian wave-packets) or continuous scattering models (such as QTBM) successfully extract resonance energies and basic decay times, they fall short in several critical aspects of transient dynamics. Specifically, the standard Landauer approach intrinsically lacks the capacity to capture explicitly time-domain  processes. Conversely, independent single Gaussian packets cannot accurately illustrate the real-time formation of a Non-Equilibrium Steady State (NESS) arising from collective multi-particle injection nor the dynamics associated with time-dependent source voltage.

Our methodology, based on the coherent sequence of orthogonal Wannier trains, successfully demonstrates the real-time formation of NESS and its application to analyzes concerning both standard characteristics obtainable from static QTBM methods and the dynamic response to the varying signal. Naturally, our approach should not be considered a universal tool applicable to every research task in the field of mesoscopic quantum transport, since, like other methods, it also suffers from limitations. Namely, although the extension to scenarios where confinement in 2D or 3D is  formally possible, the method becomes  ineffective due to computational and memory costs. However, the approach based on the explicit time evolution of each wave packet appears to be a ''sweet-spot'' when the explicit time dependent dynamics are crucial to investigate; the system can be effectively mapped onto a formally one-dimensional description.  By showing step by step the results of the numerical implementation, we prove both the validity of the approach and its potential. Namely, we have shown how NESS, representing local DC, can be constructed from the elementary wave packets. Subsequently, we have shown how typical text-book quantities such as current, shot-noise, resonant-state decay constants, or the transmission coefficient can be obtained in a reliable way. Nearly perfect agreement with QTBM based calculations allowed us to extend the analysis in a controllable manner to the  non-linear device that is the resonant tunneling diode. We have presented calculations at finite temperature and for the time-dependent carrier emission rate, which allowed us to resolve the admittance dependency on the applied bias. In this manner, we have provided \emph{proof of concept} for the method and sketched the protocol of its realization.

The usability of the packet-oriented transport modeling presented is, however, not limited to the non-interacting case. As we have also shown, this formalism can be adapted to the scenario when interaction between carriers is included in a rigorous manner. However, this approach relies on the construction of a truncated Hilbert space in which single-particle wave functions describing packets are time-independent  and are rigidly assembled based on time-injection quanta $\Delta t\sim1/\Delta E$, which provides a handy quantum number $n$ that labels orthogonal states dually with respect to the energy resolved representation. In regard to this interacting fermions scenario, the formalism can be  applied to the direct time evolution method, although it naturally does not remove the limitations concerning the small number of carriers taken into account.

However, this strategy —- by taking advantage of the second-quantization framework formulation provided by us --  maps smoothly to the modern tensor-network-based methods.

Namely, we have provided the archetypal example of transport in the regime of Coulomb blockade for the model of a quantum dot by using the matrix-product-state time evolution method. We successfully validated our outcome in this regard and believe that the proposed framework can be useful in many mesoscopic transport problems where electron-electron correlations are a crucial ingredient of the physical situation. Summarizing, the elaborated methodology allows us to study the  time-resolved dynamics of Landauer finite pulses in detail by means of analysis of both single- and many-body quantities.

In the broader landscape, we believe that the orthogonal-packets-based approach described here, which has remained a somewhat \emph{forgotten area} in the field of interest, can be efficiently exploited in a number of numerical studies concerning mesoscopic-transport. 
Furthermore, we have provided a rigorous theoretical framework that precisely connects the somewhat intuitive packet-picture with the plane-wave language natural for the L–B formalism. Therefore, besides the applicability of Wannier trains/NESS formation in the research, one may also find our explicit reasoning valuable from a didactic point of view.
Although the limitations mentioned regarding the usability of the presented approach are indisputable, the rapid development of High-Performance Computing environments may help to overcome some of the difficulties mentioned, e.g., the extension to higher spatial dimensions. We look forward to seeing progress in this matter in the near future.

\begin{acknowledgments}
This work was supported by National Science Centre (NCN) agreement number UMO-2020/38/E/ST3/00418

For the purpose of Open Access, the author has ap-
plied a CC-BY public copyright license to any Author
Accepted Manuscript (AAM) version arising from this
submission.

\end{acknowledgments}
\appendix
\section{Slicing of the energy window as a unitary transformation}
\label{app:slicing}

The slicing of the wide energy window $\Delta E$ into disjoint sub-bands (slices) $\Delta E_j$, which is utilized to significantly speed up numerical calculations, constitutes an exact unitary transformation that leaves the physical outcome invariant. To prove this, let us consider the completeness relation for the single-particle spatial states restricted to the energy window $\Delta E$. Expressing this relation in terms of the Wannier wave-packet functions $W_n(x,t)$ and expanding it back into the stationary plane waves $\psi_E(x)$, we obtain:
\begin{equation}
\begin{split}
    \sum_{n\in\mathbb{Z}} W_n(x,t) W_n^*(x',t) &= \sum_{n\in\mathbb{Z}} \left( \int_{\Delta E} \frac{1}{\sqrt{\Delta E}}e^{-iE(t-t_n)/\hbar}\psi_E(x) dE \right) \\
    &\quad \times \left( \int_{\Delta E}\frac{1}{\sqrt{\Delta E}}e^{iE'(t-t_n)/\hbar}\psi_{E'}^*(x') dE' \right) \\
    &= \frac{1}{\Delta E}\iint_{\Delta E} e^{-it(E-E')/\hbar} \\
    &\quad \times \sum_{n\in\mathbb{Z}} e^{it_n(E-E')/\hbar}\psi_E(x)\psi_{E'}^*(x')dEdE'.        
\end{split}
\end{equation}
By recognizing that $t_n=n\Delta t$ with the time-step $\Delta t=h/\Delta E$, the sum over the spatial nodes yields a Dirac comb:
\begin{equation}
    \sum_{n\in\mathbb{Z}} e^{it_n(E-E')/\hbar}=\sum_{n\in\mathbb{Z}} e^{i2\pi n(E-E')/\Delta E}=\Delta E\sum_{m\in\mathbb{Z}}\delta(E-E'+m\Delta E).
\end{equation}
Since the integrations are  bounded within a single interval of width $\Delta E$, only the Dirac delta corresponding to $m=0$ is operative. The explicit physical time $t$ thus cancels out, leading to the exact equivalence between the wave-packet and stationary energy representations of the completeness relation:
\begin{equation}
    \sum_{n \in \mathbb{Z}} W_n(x,t) W_n^*(x',t) = \int_{\Delta E}\psi_E(x)\psi_E^*(x')dE.
\end{equation}
By employing the domain additivity of the integral, one can rewrite the right-hand side of the above equation according to the subdivision of $\Delta E$ into slices $\Delta E_j$, provided that $\Delta E_i \cap \Delta E_j = \emptyset$ for $i \neq j$ and $\bigcup_{j=1}^M\Delta E_j=\Delta E$:
\begin{equation}
\begin{split}
    \int_{\Delta E}\psi_E(x)\psi_E^*(x')dE=\sum_{j=1}^{M}\int_{\Delta E_j}\psi_E(x)\psi_E^*(x')dE=&\\\sum_{j=1}^{M}\sum_{n\in \mathbb{Z}} W_{n,j}(x,t) W_{n,j}^*(x',t),
    \end{split}
\end{equation}
where we have introduced a multi-band set of Wannier packets $W_{n,j}(x,t)$, labeled by an additional quantum number $j$ denoting the specific sub-band within $\Delta E$. Consequently, we arrive at the exact spatial identity:
\begin{equation}
    \sum_{n\in{\mathbb{Z}}} W_n(x,t) W_n^*(x',t) = \sum_{j=1}^M \sum_{n\in \mathbb{Z}} W_{n,j}(x,t) W_{n,j}^*(x',t).
\end{equation}
Since the mutually disjoint energy domains guarantee the orthogonality between different sub-bands, the spatial overlap vanishes for $j \neq j'$, yielding $\int W_{n,j}^*(x,t)W_{n',j'}(x,t)dx=\delta_{nn'}\delta_{jj'}$. Thus, the partitioned set $\{W_{n,j}(x,t)\}$ constitutes a complete, mutually orthonormal basis for the entire $\Delta E$ energy window, demonstrating that the slicing procedure is an exact unitary change of basis and, in turn, does not influence the underlying physical picture but is  useful from a computational point of view.
\section{Derivation of current--current correlations and shot noise via Wick's theorem}
\label{app:wick_shot_noise}

In this section, we provide a detailed step-by-step derivation of the current--current correlation functions, Eqs.\ref{eq:sxt}-\ref{eq:sxtWick}, and the time-integrated shot noise formula, Eq.~(\ref{eq:shot_noise}), discussed in Sec.~II.D.

\subsection{ Representation and contractions}

Following Eq.~(\ref{eq:fieldopertor_decomposed}), the fermionic field operator in the Heisenberg picture is decomposed across the orthogonal direct sum $\mathcal{H}_{tot} = \mathcal{H} \oplus \mathcal{H}_q$ as
\begin{equation}
\begin{split}
    \hat{\Psi}(x,t) &= \sum_{|n\rangle \in \mathcal{H}} W_n(x,t)\hat{A}_n
    + \int_{|q\rangle \in \mathcal{H}_q} dq\,\phi_q(x,t)\hat{c}_q \\
    &\equiv \sumint_{|\mu\rangle \in \mathcal{H}_{tot}} d\mu\,\phi_\mu(x,t)\hat{\alpha}_\mu,
\end{split}
\label{eq:app_field}
\end{equation}
where $\hat{\alpha}_\mu \equiv \hat{A}_n$ for the discrete subspace $\mathcal{H}$ of injected Wannier wave packets and $\hat{\alpha}_\mu \equiv \hat{c}_q$ for the continuous subspace $\mathcal{H}_q$ of unoccupied states. Because the transformation from energy eigenstates to Wannier packets within the transport window is unitary, the unified operators $\{\hat{\alpha}_\mu, \hat{\alpha}_\nu^\dagger\}$ obey canonical fermionic anticommutation relations across the entire Hilbert space $\mathcal{H}_{tot}$:
\begin{equation}
\begin{split}
    \{\hat{\alpha}_\mu, \hat{\alpha}_\nu^\dagger\} &= \delta_{\mu\nu}, \\
    \{\hat{\alpha}_\mu, \hat{\alpha}_\nu\} &= \{\hat{\alpha}_\mu^\dagger, \hat{\alpha}_\nu^\dagger\} = 0,
\end{split}
\label{eq:app_car}
\end{equation}
where the generalized delta symbol $\delta_{\mu\nu}$ stands for the Kronecker delta $\delta_{nm}$ when $\mu, \nu \in \mathcal{H}$, the Dirac delta $\delta(q-q')$ when $\mu, \nu \in \mathcal{H}_q$, and zero when $\mu$ and $\nu$ belong to different subspaces.

In terms of $\hat{\alpha}_\mu$, the current operator reads
\begin{equation}
    \hat{I}(x,t) = e \sumint d\mu\,d\nu\, I_{\mu\nu}(x,t)\,\hat{\alpha}_\mu^\dagger \hat{\alpha}_\nu,
    \label{eq:app_I_op}
\end{equation}
with the Hermitian single-particle current matrix elements $I_{\mu\nu}(x,t) = I_{\nu\mu}^*(x,t)$ defined in Eq.~(\ref{eq:matrix_current}).

The initial many-body state $|\Phi_0\rangle = \prod_{j \in \mathcal{H}} \hat{A}_j^\dagger |0\rangle$ is a single Slater determinant in which all Wannier packet modes $|n\rangle \in \mathcal{H}$ are fully occupied, whereas all modes $|q\rangle \in \mathcal{H}_q$ are empty:
\begin{equation}
\begin{split}
    \hat{A}_n^\dagger |\Phi_0\rangle &= 0 \quad (\forall |n\rangle \in \mathcal{H}), \\
    \hat{c}_q |\Phi_0\rangle &= 0 \quad (\forall |q\rangle \in \mathcal{H}_q).
\end{split}
\end{equation}
Consequently, the elementary two-point contractions evaluated in the state $|\Phi_0\rangle$ take the diagonal projecting forms:
\begin{equation}
    \langle \Phi_0 | \hat{\alpha}_\mu^\dagger \hat{\alpha}_\nu | \Phi_0 \rangle =
    \begin{cases}
        \delta_{nm}, & |\mu\rangle = |n\rangle,\, |\nu\rangle = |m\rangle \in \mathcal{H}, \\
        0, & \text{otherwise},
    \end{cases}
    \label{eq:app_contr_1}
\end{equation}
and, by using $\hat{\alpha}_\rho \hat{\alpha}_\sigma^\dagger = \delta_{\rho\sigma} - \hat{\alpha}_\sigma^\dagger \hat{\alpha}_\rho$,
\begin{equation}
\begin{split}
    \langle \Phi_0 | \hat{\alpha}_\rho \hat{\alpha}_\sigma^\dagger | \Phi_0 \rangle
    &= \delta_{\rho\sigma} - \langle \Phi_0 | \hat{\alpha}_\sigma^\dagger \hat{\alpha}_\rho | \Phi_0 \rangle \\
    &= \begin{cases}
        \delta(q - q'), & |\rho\rangle = |q\rangle,\, |\sigma\rangle = |q'\rangle \in \mathcal{H}_q, \\
        0, & \text{otherwise},
    \end{cases}
\end{split}
\label{eq:app_contr_2}
\end{equation}
while all anomalous contractions vanish identically due to particle-number conservation, $\langle \Phi_0 | \hat{\alpha}_\mu \hat{\alpha}_\nu | \Phi_0 \rangle = \langle \Phi_0 | \hat{\alpha}_\mu^\dagger \hat{\alpha}_\nu^\dagger | \Phi_0 \rangle = 0$.

\subsection{Wick's theorem for two-time current--current correlations}

We now evaluate the two-point current fluctuation correlator
\begin{equation}
\begin{split}
    S(x,t,x',t') &\equiv \langle \Phi_0 | \Delta\hat{I}(x',t') \Delta\hat{I}(x,t) | \Phi_0 \rangle \\
    &= \langle \Phi_0 | \hat{I}(x',t') \hat{I}(x,t) | \Phi_0 \rangle - I(x',t') I(x,t),
\end{split}
\end{equation}
where $I(x,t) = \langle \Phi_0 | \hat{I}(x,t) | \Phi_0 \rangle = e \sum_{|n\rangle \in \mathcal{H}} I_{nn}(x,t)$. Substituting Eq.~(\ref{eq:app_I_op}) yields
\begin{equation}
\begin{split}
    S(x,t,x',t') = e^2 \sumint &d\mu\,d\nu\,d\rho\,d\sigma\, I_{\mu\nu}(x',t') I_{\rho\sigma}(x,t) \\
    \times \Big[ &\langle \Phi_0 | \hat{\alpha}_\mu^\dagger \hat{\alpha}_\nu \hat{\alpha}_\rho^\dagger \hat{\alpha}_\sigma | \Phi_0 \rangle \\
    &- \langle \Phi_0 | \hat{\alpha}_\mu^\dagger \hat{\alpha}_\nu | \Phi_0 \rangle
       \langle \Phi_0 | \hat{\alpha}_\rho^\dagger \hat{\alpha}_\sigma | \Phi_0 \rangle \Big].
\end{split}
\label{eq:app_S_4pt}
\end{equation}
Since $|\Phi_0\rangle$ is a single Slater determinant constructed as a direct product of fermionic creation operators in Fock space, Wick's theorem decomposes the four-operator expectation value into the sum of all complete pairwise contractions with appropriate fermionic permutation signs:
\begin{equation}
\begin{split}
    \langle \Phi_0 | \hat{\alpha}_\mu^\dagger \hat{\alpha}_\nu \hat{\alpha}_\rho^\dagger \hat{\alpha}_\sigma | \Phi_0 \rangle
    &= \langle \Phi_0 | \hat{\alpha}_\mu^\dagger \hat{\alpha}_\nu | \Phi_0 \rangle
       \langle \Phi_0 | \hat{\alpha}_\rho^\dagger \hat{\alpha}_\sigma | \Phi_0 \rangle \\
    &\quad - \langle \Phi_0 | \hat{\alpha}_\mu^\dagger \hat{\alpha}_\rho^\dagger | \Phi_0 \rangle
             \langle \Phi_0 | \hat{\alpha}_\nu \hat{\alpha}_\sigma | \Phi_0 \rangle \\
    &\quad + \langle \Phi_0 | \hat{\alpha}_\mu^\dagger \hat{\alpha}_\sigma | \Phi_0 \rangle
             \langle \Phi_0 | \hat{\alpha}_\nu \hat{\alpha}_\rho^\dagger | \Phi_0 \rangle \\
    &= \langle \Phi_0 | \hat{\alpha}_\mu^\dagger \hat{\alpha}_\nu | \Phi_0 \rangle
       \langle \Phi_0 | \hat{\alpha}_\rho^\dagger \hat{\alpha}_\sigma | \Phi_0 \rangle \\
    &\quad + \langle \Phi_0 | \hat{\alpha}_\mu^\dagger \hat{\alpha}_\sigma | \Phi_0 \rangle
             \langle \Phi_0 | \hat{\alpha}_\nu \hat{\alpha}_\rho^\dagger | \Phi_0 \rangle.
\end{split}
\label{eq:app_wick}
\end{equation}
The first (disconnected) term in Eq.~(\ref{eq:app_wick}) cancels the product of mean currents $-I(x',t')I(x,t)$ in Eq.~(\ref{eq:app_S_4pt}). Only the connected exchange contraction survives:
\begin{equation}
\begin{split}
    S(x,t,x',t') = e^2 \sumint &d\mu\,d\nu\,d\rho\,d\sigma\, I_{\mu\nu}(x',t') I_{\rho\sigma}(x,t) \\
    &\times \langle \Phi_0 | \hat{\alpha}_\mu^\dagger \hat{\alpha}_\sigma | \Phi_0 \rangle
            \langle \Phi_0 | \hat{\alpha}_\nu \hat{\alpha}_\rho^\dagger | \Phi_0 \rangle.
\end{split}
\end{equation}
Inserting the fundamental contractions (\ref{eq:app_contr_1}) and (\ref{eq:app_contr_2}), we see that $\langle \Phi_0 | \hat{\alpha}_\mu^\dagger \hat{\alpha}_\sigma | \Phi_0 \rangle$ restricts the indices $(\mu,\sigma)$ to the occupied Wannier subspace $\mathcal{H}$ with $\mu = \sigma = n$, whereas $\langle \Phi_0 | \hat{\alpha}_\nu \hat{\alpha}_\rho^\dagger | \Phi_0 \rangle$ restricts $(\nu,\rho)$ to the unoccupied subspace $\mathcal{H}_q$ with $\nu = \rho = q$. Collapsing the delta functions immediately yields Eq.~(22) of the main text:
\begin{equation}
    S(x,t,x',t') = e^2 \sum_{|n\rangle \in \mathcal{H}} \int_{|q\rangle \in \mathcal{H}_q} dq\, I_{nq}(x',t') I_{qn}(x,t).
    \label{eq:app_S_mixed}
\end{equation}

\subsection{Elimination of the unoccupied subspace $\mathcal{H}_q$ via completeness}

To express the observables solely in terms of the finite set of propagated Wannier packets $|n\rangle \in \mathcal{H}$, we invoke the completeness of the single-particle basis in $\mathcal{H}_{tot} = \mathcal{H} \oplus \mathcal{H}_q$ at any time $t$:
\begin{equation}
\begin{split}
    \sumint_{|\mu\rangle \in \mathcal{H}_{tot}} d\mu\, |\phi_\mu(t)\rangle\langle\phi_\mu(t)| &= \mathbf{I}, \\
    \int_{|q\rangle \in \mathcal{H}_q} dq\, |\phi_q(t)\rangle\langle\phi_q(t)|
    &= \mathbf{I} - \sum_{|m\rangle \in \mathcal{H}} |\phi_m(t)\rangle\langle\phi_m(t)|.
\end{split}
\label{eq:app_completeness}
\end{equation}

\paragraph{Equal-time non-local correlations ($t = t'$, $x \neq x'$).}
In first quantization, the current matrix element is $I_{\mu\nu}(x,t) = \langle \phi_\mu(t) | \hat{\mathcal{J}}(x) | \phi_\nu(t) \rangle$, where $\hat{\mathcal{J}}(x) = \frac{1}{2}\{\hat{v}_g, \delta(\hat{x}-x)\}$ is the local single-particle current density operator at point $x$. Setting $t = t'$ in Eq.~(\ref{eq:app_S_mixed}) and inserting the completeness relation (\ref{eq:app_completeness}) gives
\begin{equation}
\begin{split}
    S(x,x',t) &= e^2 \sum_{|n\rangle \in \mathcal{H}} \langle \phi_n(t) | \hat{\mathcal{J}}(x') \hat{\mathcal{J}}(x) | \phi_n(t) \rangle \\
    &\quad - e^2 \sum_{|n\rangle \in \mathcal{H}} \sum_{|m\rangle \in \mathcal{H}} I_{nm}(x',t) I_{mn}(x,t).
\end{split}
\end{equation}
For spatially separated points $x \neq x'$, the local operators $\hat{\mathcal{J}}(x')$ and $\hat{\mathcal{J}}(x)$ have disjoint spatial supports, so their operator product vanishes, $\hat{\mathcal{J}}(x')\hat{\mathcal{J}}(x) = 0$. Consequently, the single-particle contact term drops out and $S(x,x',t)$ reduces entirely to the two-particle exchange sum over the occupied Wannier states $|n\rangle, |m\rangle \in \mathcal{H}$, recovering Eq.~(\ref{eq:correlations_non_local}).

\paragraph{Time-integrated shot noise $S(x_D)$.}
Integrating Eq.~(\ref{eq:app_S_mixed}) over both time arguments $t$ and $t'$ at the detector position $x = x' = x_D$ factorizes the double time integral into the charge-transfer matrix elements $P_{\mu\nu}(x_D) = \int_{-\infty}^{\infty} I_{\mu\nu}(x_D,t)\,dt$:
\begin{equation}
\begin{split}
    S(x_D) &= \int_{-\infty}^{\infty} dt \int_{-\infty}^{\infty} dt'\, S(x_D,t,x_D,t') \\
    &= e^2 \sum_{|n\rangle \in \mathcal{H}} \int_{|q\rangle \in \mathcal{H}_q} dq\, P_{nq}(x_D) P_{qn}(x_D).
\end{split}
\label{eq:app_shot_1}
\end{equation}
From the single-particle continuity equation, $\partial_t [\phi_\mu^*(x,t)\phi_\nu(x,t)] + \partial_x I_{\mu\nu}(x,t) = 0$, and the initial condition that all wave packets originate to the left of the detector in the distant past ($\phi_\mu(x \ge x_D, t \to -\infty) = 0$), the time integral of $I_{\mu\nu}(x_D,t)$ is equal to the asymptotic matrix element of the spatial step-function projector $\hat{\Theta}_{x_D} \equiv \int_{x_D}^{\infty} dx\,|x\rangle\langle x|$:
\begin{equation}
\begin{split}
    P_{\mu\nu}(x_D) &= \int_{-\infty}^{\infty} I_{\mu\nu}(x_D,t)\,dt \\
    &= \lim_{t \to \infty} \int_{x_D}^{\infty} \phi_\mu^*(x,t)\phi_\nu(x,t)\,dx \\
    &= \lim_{t \to \infty} \langle \phi_\mu(t) | \hat{\Theta}_{x_D} | \phi_\nu(t) \rangle.
\end{split}
\label{eq:app_P_proj}
\end{equation}
Crucially, because $\hat{\Theta}_{x_D}$ is an orthogonal spatial projection operator, it is idempotent: $\hat{\Theta}_{x_D}^2 = \hat{\Theta}_{x_D}$. Using the completeness relation (\ref{eq:app_completeness}), the full sum-integral over $\mathcal{H}_{tot}$ therefore simplifies to a single diagonal matrix element:
\begin{equation}
\begin{split}
    &\sumint_{|\mu\rangle \in \mathcal{H}_{tot}} d\mu\, P_{n\mu}(x_D) P_{\mu n}(x_D) \\
    &= \lim_{t \to \infty} \langle \phi_n(t) | \hat{\Theta}_{x_D}
    \left( \sumint_{|\mu\rangle \in \mathcal{H}_{tot}} d\mu\, |\phi_\mu(t)\rangle\langle\phi_\mu(t)| \right)
    \hat{\Theta}_{x_D} | \phi_n(t) \rangle \\
    &= \lim_{t \to \infty} \langle \phi_n(t) | \hat{\Theta}_{x_D}^2 | \phi_n(t) \rangle
     = P_{nn}(x_D).
\end{split}
\label{eq:app_P_idem}
\end{equation}

Splitting the sum-integral in Eq.~(\ref{eq:app_P_idem}) into $\mathcal{H}$ and $\mathcal{H}_q$,
\begin{equation}
\begin{split}
    \int_{|q\rangle \in \mathcal{H}_q} dq\, P_{nq}(x_D) P_{qn}(x_D)
    &= \sumint_{|\mu\rangle \in \mathcal{H}_{tot}} d\mu\, P_{n\mu}(x_D) P_{\mu n}(x_D) \\
    &\quad - \sum_{|m\rangle \in \mathcal{H}} P_{nm}(x_D) P_{mn}(x_D) \\
    &= P_{nn}(x_D) - \sum_{|m\rangle \in \mathcal{H}} |P_{nm}(x_D)|^2,
\end{split}
\end{equation}
and substituting this identity into Eq.~(\ref{eq:app_shot_1}) yields the final shot-noise formula, Eq.~(\ref{eq:shot_noise}) of the main text:
\begin{equation}
    S(x_D) = e^2 \sum_{|n\rangle \in \mathcal{H}} \left( P_{nn}(x_D) - \sum_{|m\rangle \in \mathcal{H}} |P_{nm}(x_D)|^2 \right),
\end{equation}
which depends exclusively on the occupied Wannier wave packets $|n\rangle \in \mathcal{H}$.

\section{Matrix product state architecture}
\label{sec:appB}
As discussed in the main text, mapping the quantum transport problem onto the
rigid-Wannier basis within an MPS framework requires tailoring the 1D chain
topology to minimize long-range entanglement and keep the maximum bond
dimension $\chi_{\max}$ under control. In our truncated three-band setup
($M = 100$ positions in $\epsilon_t$ and $2K_{\mathrm{virt}} + 1 = 21$
positions $n \in [n_c - K_{\mathrm{virt}}, n_c + K_{\mathrm{virt}}]$ around
$n_c = M/2 = 50$ in $\epsilon_b$ and $\epsilon_u$), the $284$ spatial orbitals
($\mathcal{N} = 568$ spinful orbitals) are arranged along the 1D MPS chain
$\mathcal{C} = \mathcal{R}_{\text{in}} \oplus \mathcal{R}_{\text{QD}} \oplus \mathcal{R}_{\text{out}}$
as follows:
\begin{enumerate}
    \item \textbf{Upstream incoming and background reservoir ($\mathcal{R}_{\text{in}}$):}
    Since the spatial centers of the rigid Wannier packets are given by
    $x_c^{n,\zeta} \approx \zeta(n - n_c)\Delta x_j$, the left-moving transport
    modes with $n < n_c - 1$ ($\zeta = -1$, located at $x_c^{n,-1} > 0$)
    propagate from the drain towards the barrier and remain virtually inert in
    the absence of right-side injection. We therefore place the
    $(n \in [0, n_c - 2], \epsilon_t, -1)$ modes at the far-left boundary of
    the MPS chain (positions $1\text{--}49$, where $\chi \approx 1$), followed
    by the outer virtual-band dressing orbitals with
    $2 \leq |n - n_c| \leq K_{\mathrm{virt}}$ from the upper ($\epsilon_u$,
    $36$ orbitals) and bottom ($\epsilon_b$, $36$ orbitals) sub-bands, and
    subsequently the incident right-moving transport channel
    $(n \in [0, n_c - 2], \epsilon_t, +1)$ ordered monotonically towards the
    barrier ($n = 0, 1, \dots, n_c - 2$). In this manner, the initial
    $N$-fermion Landauer train ($N = 2N_{\uparrow} \in \{4, 8, 12\}$) occupying
    $n \in [n_c - 1 - N_{\uparrow}, n_c - 2]$ in the conduction sub-band
    $\epsilon_t$ is initialized at $t = 0$ as an exact product state of unit
    bond dimension ($\chi = 1$),
    \begin{equation}
        |\Phi_0\rangle = \prod_{n = n_c - 1 - N_{\uparrow}}^{n_c - 2}
        \hat{c}_{n,\epsilon_t,+1,\uparrow}^{\dagger}\,
        \hat{c}_{n,\epsilon_t,+1,\downarrow}^{\dagger}|0\rangle,
    \end{equation}
    situated directly adjacent to the central interaction core.

    \item \textbf{Quantum-dot interaction core and barrier-slope bridge ($\mathcal{R}_{\text{QD}}$):}
    The central block $\mathcal{R}_{\text{QD}}$ groups together the $18$
    spatial orbitals ($36$ spinful orbitals) belonging to the three core
    positions $n \in \{n_c - 1, n_c, n_c + 1\} = \{49, 50, 51\}$ across all
    three sub-bands $j \in \{\epsilon_b, \epsilon_t, \epsilon_u\}$ and both
    branches $\zeta = \pm 1$. Within this $18$-orbital core block, we retain
    the complete four-index off-diagonal Coulomb interaction tensor
    $\mathcal{V}_{pqrs}\,\hat{c}_{p\sigma}^{\dagger}\hat{c}_{q\sigma'}^{\dagger}\hat{c}_{s\sigma'}\hat{c}_{r\sigma}$
    (including interband, correlated-hopping, and pair-exchange processes),
    supplemented by the two-index density-density Coulomb interaction
    $U_{km}\hat{N}_k\hat{N}_m$ for orbital pairs extending outside the core.

    Strict attention is paid to the dot center $n = n_c = 50$ ($x = 0$), where
    the right- and left-moving Wannier functions are complex conjugates sharing
    an identical spatial envelope,
    $W_{n_c,j,-1}(x) = W_{n_c,j,+1}^{*}(x) = W_{n_c,j,+1}(-x)$, and are
    strongly mixed by the symmetric double-barrier potential
    $V_{\mathrm{QD}}(-x) = V_{\mathrm{QD}}(x)$. Retaining the chiral basis
    $(\zeta = \pm 1)$ at $n = n_c$ would generate strong artificial intra-dot
    entanglement across the central MPS bonds. To eliminate this bottleneck, we
    perform a local orthogonal transformation at $n = n_c$ into even ($+$) and
    odd ($-$) spatial parity Wannier states for each sub-band
    $j \in \{\epsilon_b, \epsilon_t, \epsilon_u\}$:
    \begin{equation}
        \hat{c}_{n_c,j,\pm,\sigma}^{\dagger} =
        \frac{1}{\sqrt{2}}\Big(\hat{c}_{n_c,j,+1,\sigma}^{\dagger}
        \pm \hat{c}_{n_c,j,-1,\sigma}^{\dagger}\Big),
    \end{equation}
    and correspondingly rotate the one-body Hamiltonian and the four-index core
    Coulomb tensor $\mathcal{V}_{pqrs}$ into the parity-adapted basis.
    Crucially, because $|W_{n_c,j,+1}(x)|^2 = |W_{n_c,j,-1}(x)|^2$, the
    intra-dot density-density Coulomb matrix elements are identical for both
    chiralities, rendering the diagonal local interaction in each sub-band
    invariant under the $(+1, -1) \to (+, -)$ rotation:
    \begin{equation}
    \begin{split}
        \hat{\mathcal{U}}_{n_c,j} = \frac{\mathcal{V}_{n_c j, n_c j}}{2}\,
        \hat{N}_{n_c,j}\big(\hat{N}_{n_c,j} - 1\big), \\
        \hat{N}_{n_c,j} = \sum_{\alpha \in \{+,-\}}
        \big(\hat{n}_{n_c,j,\alpha,\uparrow} + \hat{n}_{n_c,j,\alpha,\downarrow}\big).
        \end{split}
    \end{equation}
    Along the MPS segment $\mathcal{R}_{\text{QD}}$, the six upstream
    barrier-slope orbitals at $n = n_c - 1$ (ending with the incident mode
    $(n_c - 1, \epsilon_t, +1)$) are placed immediately before the six
    parity-adapted dot modes
    $\big[(n_c, \epsilon_t, +), (n_c, \epsilon_b, +), (n_c, \epsilon_u, +),
    (n_c, \epsilon_u, -), \\(n_c, \epsilon_b, -), (n_c, \epsilon_t, -)\big]$,
    followed by the six downstream barrier-slope orbitals at $n = n_c + 1$.
    This compact $18$-orbital bridge captures both the resonant even-parity
    charge accumulation inside the well and the off-resonant hybridization with
    the barrier slopes while keeping all four-fermion interaction operators
     local in the MPS chain.

    \item \textbf{Downstream outgoing channels ($\mathcal{R}_{\text{out}}$):}
    For $n > n_c + 1$ ($n = n_c + 2, \dots, M - 1$), both the transmitted
    right-moving packets $(n, \epsilon_t, +1)$ (propagating into $x > 0$) and
    the reflected left-moving packets $(n, \epsilon_t, -1)$ (propagating back
    into $x < 0$) move radially away from the QD with the same temporal delay
    $(n - n_c)\Delta t_{\epsilon_t}$. Since each partial scattering event at
    the barrier entangles transmitted and reflected components of equal radial
    distance $(n - n_c)$, we interleave them pairwise along the MPS chain as
    $\big[(n, \epsilon_t, +1), (n, \epsilon_t, -1)\big]_{n=n_c+2}^{M-1}$,
    keeping the mutually entangled outgoing modes  local.
\end{enumerate}

\bibliography{bibliography}

\end{document}